\documentclass[fleqn,usenatbib]{mnras}

\usepackage{newtxtext,newtxmath}

\usepackage[T1]{fontenc}

\DeclareRobustCommand{\VAN}[3]{#2}
\let\VANthebibliography\thebibliography
\def\thebibliography{\DeclareRobustCommand{\VAN}[3]{##3}\VANthebibliography}

\usepackage{graphicx}	
\usepackage{amsmath}	

\pdfoutput=1
\usepackage[T1]{fontenc}
\usepackage{graphicx}
\usepackage{wasysym}
\usepackage{esint}
\usepackage[authoryear]{natbib}
\usepackage[utf8]{inputenc}
\usepackage{threeparttable}
\usepackage{hyperref}
\usepackage{xspace}
\usepackage{url}
\usepackage{babel}
\usepackage{ulem}
\usepackage{newtxtext,newtxmath}
\usepackage{hyperref}	
\usepackage{mathtools}
\usepackage{xcolor}
\usepackage{multirow}
\usepackage{orcidlink}
\usepackage{pifont}

\newcommand{\cmark}{\ding{51}} 
\newcommand{\xmark}{\ding{55}} 

\hypersetup{colorlinks=true,linkcolor=blue,citecolor=blue,filecolor=blue,urlcolor=blue,}

\newcommand{\arepo}{\textsc{Arepo}\xspace}

\newcommand{\dd}{\mathrm{d}}

\newcommand{\msun}{\mathrm{M}_\odot}

\newcommand{\crayon}{\textsc{Crayon+}\xspace}
\newcommand{\crest}{\textsc{Crest}\xspace}
\newcommand{\B}{{\mathcal B}}

\newcommand{\e}{\mathrm{e}}

\title[\crest in galaxies]{
Steady-State or Not? The Evolution of Cosmic Ray Electron Spectra in Galaxies
}
\author[M. Werhahn et al.]{Maria Werhahn$^{1,}$\thanks{E-mail:
mwerhahn@mpa-garching.mpg.de}\orcidlink{0000-0003-4984-4389}, 
Christoph Pfrommer$^2$\orcidlink{0000-0002-7275-3998},
Joseph Whittingham$^2$\orcidlink{0000-0002-0198-8490},
L\'{e}na Jlassi$^2$\orcidlink{0009-0007-9039-294X},
R\"udiger Pakmor$^1$\orcidlink{0000-0003-3308-2420},
\newauthor Philipp Girichidis$^3$\orcidlink{0000-0002-9300-9914}, 
Rebekka Bieri$^4$\orcidlink{0000-0002-4554-4488}
\\
$^{1}$Max-Planck-Institut f\"ur Astrophysik (MPA), Karl-Schwarzschild-Str. 1, 85748 Garching, Germany\\
$^{2}$Leibniz-Institut f\"{u}r Astrophysik Potsdam (AIP), 
    An der Sternwarte 16, D-14482 Potsdam, Germany\\
$^{3}$Universit\"{a}t Heidelberg, Zentrum f\"{u}r Astronomie, Institut f\"{u}r Theoretische Astrophysik, Albert-Ueberle-Str. 2, 69120 Heidelberg, Germany\\
$^{4}$Department of Astrophysics, University of Zurich, 8057 Zurich, Switzerland\\%
}

\date{Accepted XXX. Received YYY; in original form ZZZ}

\pubyear{2024}

\begin{document}
\label{firstpage}
\pagerange{\pageref{firstpage}--\pageref{lastpage}}
\maketitle

\begin{abstract} 
Cosmic ray (CR) electrons are key tracers of non-thermal processes in galaxies, yet their spectra are often modelled under the untested assumption of steady state between injection and cooling. In this work, we present a time-dependent modelling of CR electron spectra in a galactic context using the \crest code, applied to magnetohydrodynamical simulations of an isolated Milky Way-mass galaxy performed with \arepo. CR electrons are injected at supernova sites and evolved with adiabatic changes and cooling processes on Lagrangian tracer particles, including losses from synchrotron, inverse Compton, bremsstrahlung, and Coulomb interactions. We compare these fully time-dependent spectra to local and global steady-state models computed with \crayon, as well as to one-zone analytic steady-state solutions.
We find that the global CR electron spectrum in the simulated galactic disk closely resembles a steady-state solution up to energies of 500~GeV, with deviations only at higher energies where cooling times become shorter than injection timescales. High-energy electrons are dominated by recently injected populations that have not yet reached equilibrium, however, producing a steeper spectrum and lower normalisation than a steady-state model predicts.
Spatially, the electrons modelled on-the-fly with \crest are more confined to the star-forming disk, in contrast to the more extended distributions from steady-state post-processing models. Our results demonstrate that while steady-state assumptions capture the bulk CR electron population in star-forming disks, a time-dependent treatment is essential to describe the high-energy and outflowing components.
 \end{abstract}




\section{Introduction}

Cosmic rays (CRs) are thought to play a key role in feedback processes that regulate galaxy evolution \citep[see][for a recent review]{2023RuszkowskiPfrommer_Review}. Among them, CR protons dominate the energy budget and are therefore dynamically the most important. However, CR electrons provide a more direct window into the CR population because they are much more readily observed.
Gamma-ray emission from CR protons arises through neutral pion decay, but robust detections with current observational facilities are limited to the Milky Way (MW) and a handful of nearby star-forming galaxies. 

In contrast, CR electrons, despite their lower overall energy density, are better observable through the synchrotron radiation they produce in the presence of magnetic fields. This emission is typically detected in the radio continuum and has been extensively studied across large samples of nearby galaxies. Observations reveal, for example, a tight correlation between radio luminosity and star formation rate \citep[SFR;][]{1971VanDerKruit, 1985Helou, 1992Condon, Yun_2001, 2003Bell, 2021Molnar, 2021Matthews}, detailed radio spectra that may constrain the underlying CR electron distributions \citep[e.g.][]{2011Strong, 2016Mulcahy}, 
and large radio halos in edge-on systems that trace galactic outflows on kiloparsec scales both in normal star-forming galaxies \citep[e.g.][]{2020Stein,2024Heesen,2025Matthews} as well as in galaxies with a potential active galactic nucleus \citep{2025Veronese}.

Together, these observational insights highlight the importance of modelling CR electron spectra in a full galactic context. Such modelling not only advances our understanding of CR physics but also provides a crucial bridge between theory and the wealth of radio data available for star-forming galaxies.

A recent advance in this direction has been achieved by \citet{2021Ogrodnik}, who implemented a method for momentum-dependent propagation of CR electrons in MHD simulations using the \textsc{Piernik} grid code. They validated their approach in a stratified-box setup of a CR-driven galactic wind, demonstrating the feasibility of following the spectral evolution of CR electrons in self-consistent MHD simulations.
Furthermore, \citet{2024aPonnadaSpectrallyResolved} presented spectrally resolved predictions of synchrotron emission from MHD galaxy formation simulations within the FIRE-2 framework. Their simulations evolve both magnetic fields and CR proton and electron spectra \citep[introduced in][]{2022Hopkins} using a constant, empirically calibrated scattering rate, demonstrating that explicit spectral evolution is crucial for accurate synchrotron predictions, particularly in dense galactic centres where energy losses are strong. In a complementary study, \citet{2024bPonnada} investigated the impact of different CR proton transport models based on self-confinement and extrinsic turbulence. However, they adopted a fixed proton-to-electron ratio and a constant electron spectral shape, thereby neglecting spatial variations in CR losses and electron spectra that arise from differences in local gas conditions and transport physics.

Recent work by \citet{2025Linzer} has modelled the transport and spectral evolution of CR electrons in a kpc-sized TIGRESS simulation patch representative of the solar neighbourhood. Their approach couples realistic, multiphase ISM conditions with a two-moment CR transport model that includes advection, streaming, and diffusion, the latter being determined self-consistently from local gas and CR properties by balancing wave growth from the streaming instability against various damping processes. By post-processing the MHD simulations for different CR electron energies, they resolve energy-dependent cooling and transport effects for CR electrons (and protons), but do not account for momentum-space transport, which precludes a comparison of the electron spectrum with local data from Voyager and AMS-02. This model provides detailed insight into CR propagation and diffusion under realistic ISM conditions but, owing to the limited spatial domain, it cannot capture the global evolution or large-scale coupling of CR populations to the galactic structure.

Complementary to such local, high-resolution studies, large-scale Galactic propagation models such as \textsc{Galprop}, \textsc{Dragon}, \textsc{Picard}, and \textsc{Usine} \citep[e.g.][]{1998StrongMoskalenko, 2014Kissmann, 2018Johannesson, 2020Maurin} solve the CR transport equation on fixed grids, adopting parameterized source distributions and transport coefficients that are tuned to reproduce a broad range of observational constraints, including CR nuclei and lepton spectra, secondary-to-primary ratios, diffuse gamma-ray emission, and Galactic synchrotron radiation. Furthermore, one-zone leaky-box models have been developed to interpret the non-thermal emission of star-forming galaxies, treating the average gas density, magnetic field strength, and CR confinement time as adjustable parameters \citep[e.g.\ ][]{2010Lacki, 2019Peretti}. One-dimensional transport models have further been used to explain the vertical radio emission profiles of galactic haloes and to constrain the diffusion and advection parameters governing CR transport \citep{2016Heesen, 2018Heesen}.

In this work, we extend our previous studies that modelled the CR electron and proton distributions in global CR-MHD simulations and compared the resulting spectra to CR data \citep{2021WerhahnI}, $\gamma$-ray spectra, emission maps, and correlations of the SFR with the $\gamma$-ray emission \citep{2021WerhahnII,Werhahn2023} as well as radio spectra, emission maps, and the far-infrared--radio relation \citep{2021WerhahnIII,2022Pfrommer,2024ChiuSandy} while assuming the steady-state approximation in post-processing. To test the validity of this approximation for CR electrons in galaxies, we apply the \crest (Cosmic Ray Electron Spectra that are evolved in Time) code introduced in \citet{2019Winner} for the first time to simulations of galaxies. It has been already successfully applied to simulations of Sedov explosions to explain the non-thermal emission maps and profiles of SN~1006 \citep{2020Winner}, as well as to simulations of radio relics in galaxy clusters \citep{2024Whittingham}.
Here, we follow the time-dependent, spectrally-resolved evolution of CR electrons in a full galactic simulation, and we compare those results to cell-based and one-zone steady-state models and to observational data. 

The paper is organised as follows. In Section~\ref{Sec:Simulations}, we describe the numerical setup and physical ingredients of the simulations, and introduce our CR modelling methods.
In Section~\ref{Sec:Steady-state vs. CREST}, we compare spatial distributions, global and regional spectra, and temporal evolution between the time-dependent and steady-state approaches and compare them with observations from Voyager~1, AMS-02, and H.E.S.S.
In Section~\ref{sec:Memory}, we quantify the effective memory of the CR electron population by restarting the spectral simulations at different look-back times and by analysing cooling timescales. 
Finally, we summarise our main findings and outline future work in Section~\ref{Sec:Conclusion}. Additional tests and numerical checks are presented in the Appendix.

\section{Simulations and methods}
\label{Sec:Simulations}

Our simulations are performed with the moving mesh code \arepo \citep{2010Springel, 2016cPakmor}. 
Our simulation setup, apart from the additional refinement described below, is identical to that used in \citet{2017bPfrommer} and \citet{2021WerhahnI}. We initialize a gas cloud in approximate hydrostatic equilibrium and include radiative cooling and star formation following the subgrid model of \citet{2003SpringelHernquist}. The collapse of the gas cloud leads to the formation of a rotationally supported disc, accompanied by an initial peak in SFR of $62~\mathrm{M_\odot,yr^{-1}}$ at $t \approx 170~\mathrm{Myr}$. The SFR subsequently declines over time, reaching $2.9~\mathrm{M_\odot,yr^{-1}}$ at $t = 4~\mathrm{Gyr}$, by which point a total stellar mass of $4.6\times10^{10}~\mathrm{M_\odot}$ has formed. 
As a result, the system naturally evolves into a Milky Way–like disc galaxy in terms of its stellar mass, total mass and star formation rate.

It is embedded in a dark matter halo following an NFW \citep{1997Navarro} profile with a total mass of $M_{200}=10^{12}~\msun$ and a concentration parameter of $c_{200}=7$. We perform our simulations both with medium (`1M') and high resolution (`10M') variants. For the high-resolution simulation, the target mass of the initially $10^7$ gas cells is $1.55\times10^{4}~\msun$, whereas it is $1.55\times10^{5}~\msun$ for the medium resolution with $10^6$ gas cells. This mass is kept within a factor of two by refining and de-refining mesh cells. 
In addition to this, in order to better resolve the emerging galactic outflows, we run a high-resolution simulation with super-Lagrangian refinement in the central region (`10M\_refinement'). For this, following the same procedure as in \citet{2025ThomasPfrommerPakmor}, we define four nested shells around the centre of the box with $r_\mathrm{shell}=\{15, 30, 60, 120\}$~kpc. Within each of these shells, a maximum cell size of $\Delta x_\mathrm{max}=\{100, 200, 400, 1000\}$~pc is enforced by refining gas cells with volumes $V>4\pi/3 \Delta x_\mathrm{max}^3$.
We adopt this as our fiducial setup for all analyses presented in this paper, unless stated otherwise.

\begin{figure*}
    \centering
    \includegraphics[]{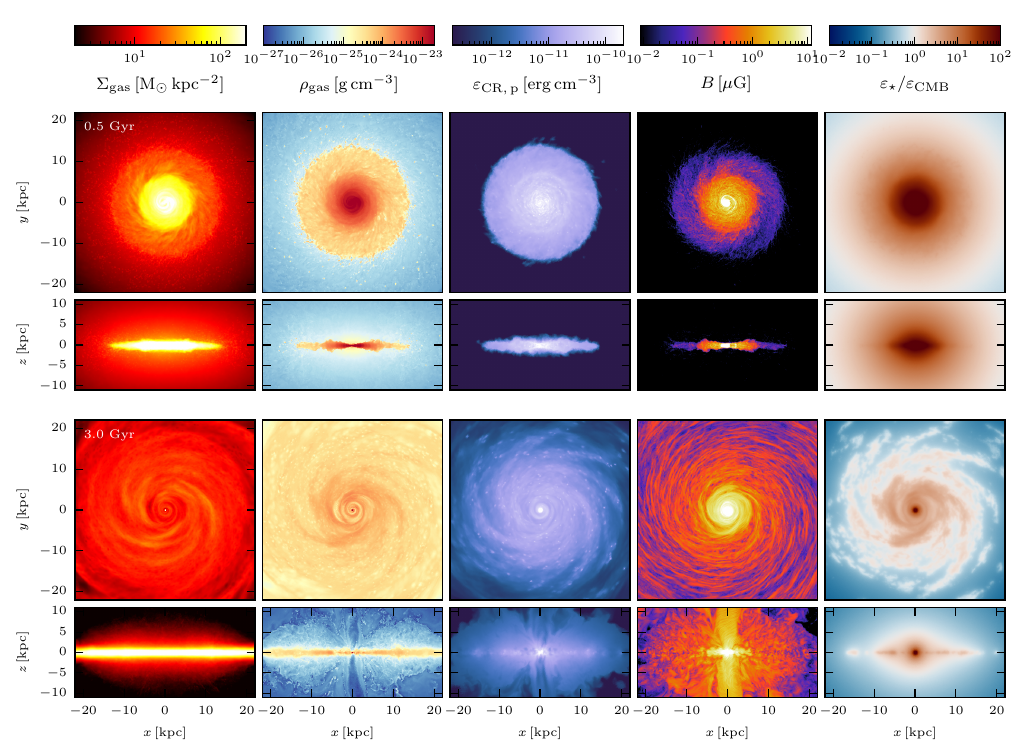}
    \caption{Overview of the simulated galaxy at 0.5~Gyrs (first two rows) and 3.0~Gyrs (last two rows) of evolution, both face-on (first and third row) and edge-on (second and fourth row). From left to right, we show the gas surface density, gas density, CR proton energy density, magnetic field strength and ratio of the interstellar radiation field $\varepsilon_\star$ to the CMB energy densities. The gas surface density is projected along 44~kpc and the magnetic field strength is averaged over a thin slice of thickness 1~kpc, overlaid with a line integral convolution indicating its orientation \citep{1993CabralLeedom}. All other quantities are shown as thin slices through the centre.} 
    \label{fig:map}
\end{figure*}

In Fig.~\ref{fig:map}, we overview physical properties at two evolutionary stages of our fiducial simulation (`10M\_refinement'). At 500~Myr (upper panels), we see the gas disk that forms after the initial collapse of the gas cloud.

We model the ISM with an effective equation of state and stochastic star formation above a critical density threshold $n_\mathrm{SF}=0.13\,\mathrm{cm^{-3}}$ \citep{2003SpringelHernquist}. Furthermore, we evolve the equations of ideal MHD \citep{2013Pakmor} with an initial seed magnetic field of strength $10^{-10}$~G. This field is exponentially amplified through adiabatic compression and a small-scale turbulent dynamo, reaching approximate equipartition with the turbulent energy density after approximately 200~Myr. At later times, the magnetic field coherence length increases progressively, extending to larger spatial scales \citep{2022Pfrommer}.

We inject a fraction of $\zeta_\mathrm{SN}=0.10$ of the kinetic energy of supernovae (SNe) into CRs. 
The CR energy from each newly formed star particle is distributed to the surrounding gas using the standard cubic spline kernel (as in SPH), normalized over the nearest $\sim$32 mesh cells within the star’s smoothing length.
The total injected energy averages out to $\Delta E_\mathrm{CR}=\zeta_\mathrm{SN}\epsilon_\mathrm{SN}\dot{m}_\star \Delta t$, where $\dot{m}_\star$ denotes the SFR of the cell and the SN energy per unit mass is given by $\epsilon_\mathrm{SN}=10^{49}\mathrm{erg\,\msun^{-1}}$.
CRs are transported in the diffusion-advection approximation with a one-moment CR hydrodynamics algorithm \citep{2016aPakmor, 2017aPfrommer}. We adopt a spatially constant diffusion coefficient of $\kappa_\mathrm{\parallel}=10^{28}\,\mathrm{cm^2\,s^{-1}}$ along the magnetic field vector. 
Hence, CR protons are not only advected with the gas in the disk but continue to diffuse above and below the disk, leading to large scale galactic outflows, as has been already found in earlier work with similar setups \citep{2016bPakmor_winds, 2018Jacob}. These outflows typically start at around 0.6~Gyr of evolution, and are still clearly visible at 3~Gyr as low-density, highly magnetised winds (see lower panels of Fig.~\ref{fig:map}).

In this paper, we present the first results of our new method of calculating the CR electron spectra on Lagrangian tracer particles with \crest in a full galaxy simulation. To benchmark our results, we compare them to a cell-based steady-state approach (using \crayon, see Section~\ref{sec:crayon}), as well as a simple one-zone steady-state model.
The \crayon\ framework has previously been applied to simulations of galaxies ranging from dwarf to MW masses, successfully reproducing the observed scaling relations and non-thermal spectra of star-forming galaxies across the radio and $\gamma$-ray regimes \citep{2021WerhahnII, 2021WerhahnIII, 2022Pfrommer, 2024ChiuSandy}.

\subsection{Live CR electron modelling with \crest}
Since CR electrons constitute only a small fraction of the total CR energy density, they do not significantly influence the dynamical evolution of the galaxy. This justifies evaluating their spectral evolution in post-processing on Lagrangian tracer particles that follow the hydrodynamic flow.
These track all relevant physical quantities on-the-fly that are required to solve the Fokker-Planck equation in post-processing using the \crest code \citep{2019Winner}. This includes adiabatic changes, as well as losses due to Coulomb interactions and losses arising from bremsstrahlung, synchrotron and inverse Compton (IC) radiation.

For Coulomb, bremsstrahlung and synchrotron cooling we can simply use the gas properties (i.e. the gas density and magnetic field strength) of the host cells of a tracer particle. For calculating IC cooling, however, we need an additional model for the incident radiation field, since this is not calculated explicitly within the simulation. As an estimate of this, and to be consistent with the post-processing steady-state code that we compare our results to, we use the same approach as in \citet{2021WerhahnI}. Specifically, we calculate the photon energy density $\varepsilon_\mathrm{ph}=\varepsilon_\mathrm{CMB}+\varepsilon_\star$, i.e.\ as the sum of the photon energy density of the CMB today ($\varepsilon_\mathrm{CMB}=4.16\times10^{-13}\, \mathrm{erg\,cm^{-3}}$) and an estimate for the interstellar radiation field $\varepsilon_\star$. The latter is calculated for each cell by summing over the far-infrared (FIR) flux $\Sigma_i \varepsilon_\star=L_\mathrm{FIR}/(4\pi R_i^2c)$ from all other cells $i$ with a non-zero SFR at distance $R_i$. The FIR luminosity from a cell is estimated from the \citet{1998Kennicutt} relation, i.e.\ $L_\mathrm{FIR}/L_\odot=1/(1.7\epsilon)\times10^{10}\times\dot{M}_\star/(\mathrm{M_\odot\,yr^{-1}})$, where the parameter $\epsilon=0.79$ is adopted assuming a \citet{2003Chabrier}
initial mass function \citep{2010Crain}. The resulting ratio of stellar to CMB energy density is shown in the right-hand column of Fig.~\ref{fig:map}. At early times, $\varepsilon_\star$ can exceed the CMB radiation field by up to a factor of 100 in the central few kpc of the disk. It decreases with time according to the decreasing SFR in the simulation.

We first run the simulation without tracer particles until a gas disc has formed ($t=500$~Myr). We then place tracer particles in each gas cell within a cylinder ($R=30$~kpc, $\lvert z\lvert < 10$~kpc) and continue the simulation from thereon including the tracer particles. This enables us to more specifically target the gas that we are interested in, due to the following limitations.
First, our Lagrangian tracer particles (also called ‘velocity field tracers’) do not perfectly follow the mass flow of the gas \citep{2013Genel}. Because their velocities are calculated from the linearly interpolated gas velocity field, they inherently cannot fully capture converging or diverging flows that occur in our simulation; in particular, during the initial collapse of the gas cloud. 
Secondly, the gas distribution changes rapidly over the first 500 Myr, as a large fraction turns into stars during the initial starburst-phase of the simulation, and the remaining cells undergo strong refinement. In contrast, tracers do not refine, which hinders their ability to follow the gas flow during this phase.
Therefore, we place the tracer particles in the simulation after a disc has formed and at least 300~Myrs before the time of analysis.
Due to the short cooling timescales of CR electrons, this allows us to still recover the full correct electron spectra, with a relative error in the total spectrum of <2.2\% (<0.16\%) after 300~Myr (500~Myr) of evolution, as we will demonstrate in Section~\ref{sec:Memory}.

This strategy of placing one tracer particle per cell when restarting the simulation from $\gtrsim$300~Myr before the time of interest leads, however, to an under-sampling of the outflows from the simulated galaxy, which are typically underdense and hence often only sparsely populated with tracer particles. To improve the sampling, we spawn additional tracer particles in the central region (defined by a cylinder of radius 5~kpc and total height of 20~kpc) in cells that are devoid of tracer (e.g.\ due to refinement of cells). These newly spawned tracers track the ID of their closest neighbouring tracer particle. In \crest, these new tracers then inherit the spectrum from their nearest neighbour and are evolved independently from thereon.

The distribution function of CR electrons is evolved in one-dimensional momentum space with $f=4\pi p^2 f^\mathrm{3D}$, where $f^\mathrm{3D}$ is the three-dimensional distribution function. We sample the CR electron spectrum from $p=10^{-1}$ to $10^8$ with 20 bins per decade, where $p=P/(m_\mathrm{e}c)$ denotes the normalised momentum, $m_\mathrm{e}$ is the electron rest mass, and $c$ is the speed of light. 
Initially, all tracer particles are assumed to have a thermal spectrum only. 

We inject CR electrons from SN, using a subgrid model for the unresolved SN remnants in our simulation.
The electrons gain a fraction $\zeta_\mathrm{ep}$ of the energy injected into CR protons $\Delta E_\mathrm{CR}=\zeta_\mathrm{SN}\epsilon_\mathrm{SN}\dot{m}_\star \Delta t$ (see Section~\ref{Sec:Simulations}).
Hence, the total injected electron energy is defined as
\begin{align}
\Delta E_\mathrm{CRe}=\zeta_\mathrm{ep}\Delta E_\mathrm{CR}.
\label{eq:zeta_pe}
\end{align}
We will discuss the choice of $\zeta_\mathrm{ep}$ in Section~\ref{sec:Normalisation}.

The source function for evolving the CR electron spectrum is given by a power law
\begin{align}
    q(p) = \frac{\tilde{C}_\mathrm{inj}}{\Delta t}\,p^{-\alpha_\mathrm{inj}} \theta(p-p_\mathrm{min}),
    \label{eq:q(p)}
\end{align}
where we set the injection slope to be $\alpha_\mathrm{inj}=2.2$.
The minimum momentum of injection, $p_\mathrm{min}$, is set by 
\begin{align}
    \int_{p_\mathrm{min}}^{\infty} q(p) E_\mathrm{kin}(p)\dd p= \frac{\Delta \varepsilon_\mathrm{CRe}}{\Delta t}
\end{align}
with the CR electron energy density $\varepsilon_\mathrm{CRe}=E_\mathrm{CRe}/V_\mathrm{cell}$, where $V_\mathrm{cell}$ is the volume of the gas cell.
To determine the normalisation $\tilde{C}_\mathrm{inj}$ of the source function, we require to match the thermal spectrum at $p_\mathrm{min}$, i.e.
\begin{align}  
    \tilde{C}_\mathrm{inj}= f_\mathrm{th}(p_\mathrm{min}) p_\mathrm{min}^{\alpha_\mathrm{inj}}.
\end{align}

Finally, we apply a modified, exponential cut-off at high energies via
\begin{align}
    \tilde{q}(p) = q(p) \times [1+ a_1(p/p_\mathrm{cut})^{a_2}]^{a_3} \times[\exp[-(p/p_\mathrm{cut})^2],
    \label{eq:tilde_q(p)}
\end{align}
with the parameters $a_1=0.66$, $a_2=2.5$, and $a_3=1.8$ \citep{2007Zirakashvili}. We set the cut-off momentum $p_\mathrm{cut}=3.9\times 10^{7}$, which corresponds to an electron energy of 20~TeV. This is motivated by X-ray observations of SNRs which suggest maximum electron energies in the range of 1-100~TeV \citep{2012Vink}.

\subsection{Steady-state CR electron modelling with \crayon}\label{sec:crayon}

In addition, we calculate steady-state spectra of CR electrons with \crayon \citep[see ][ for more details]{2021WerhahnI}. We solve the steady-state equation in each computational cell, accounting for the relevant loss terms:
\begin{equation}
\frac{\mathrm{}f(E)}{\tau_{\mathrm{esc}}}-\frac{\mathrm{d}}{\mathrm{d}E}\left[f(E)b(E)\right]=q(E),
\label{eq:diff-loss-equ}
\end{equation}
where the loss rate $b(E)$ includes synchrotron, IC, bremsstrahlung and Coulomb losses.
As a source term, we assume the same functional form as in Eq.~\eqref{eq:tilde_q(p)}. The steady-state equation also accounts for escape losses, in the form of advection and diffusion ($\tau_{\mathrm{esc}}=1/(\tau_{\mathrm{adv}}^{-1} + \tau_{\mathrm{diff}}^{-1})$). Since \crest does not account for spatial diffusion, we run \crayon both with and without diffusion and test the effect on the resulting spectra. In practice this means we set $\tau_\mathrm{diff}\gg\tau_\mathrm{Hubble}$.

\subsection{Normalisation of the electron spectra}\label{sec:Normalisation}

The only direct constraint on the electron-to-proton ratio $\zeta_\mathrm{ep}$ comes from direct measurements of the CR electron and proton spectra in the solar neighbourhood. In particular, the ratio of electrons to protons at 10~GeV \citep[$K_\mathrm{ep}^\mathrm{obs}\approx 0.01$; ][]{Cummings2016} is often used to calibrate CR electron models.
We normalise our steady-state modelling of CR electrons with \crayon by defining an injected electron-to-proton ratio ($K_\mathrm{ep}^\mathrm{inj}$), such that the cooled spectra match the observed value at 10~GeV. In the fiducial steady-state modelling including advection and diffusion, this amounts to $K_\mathrm{ep}^\mathrm{inj}\approx0.02$ \citep{2021WerhahnI}. Neglecting diffusion for either protons and/or electrons leads to small variations in this value (ranging from 0.014 to 0.049, see Table~\ref{tab:kep}).

In \crest, the normalisation is determined by the energy ratio of electrons to protons $\zeta_\mathrm{ep}$, see Eq.~\eqref{eq:zeta_pe}. It is related to $K_\mathrm{ep}^\mathrm{inj}$ via
\begin{align}
    K_\mathrm{ep}^\mathrm{inj} = \zeta_\mathrm{ep}\left(\frac{m_\mathrm{p}}{m_\mathrm{e}}\right)^{2-\alpha_\mathrm{p}}.
    \label{eq:Kep_inj,zeta_ep}
\end{align}
Note that this relation only holds if there is no cut-off in the CR spectra. Accounting for a cut-off at large momenta changes this value by $\sim$10 per cent \citep{2024ChiuSandy} for $\alpha_\mathrm{inj}=2.2$.
Following Eq.~\ref{eq:Kep_inj,zeta_ep}, a value of $K_\mathrm{ep}^\mathrm{inj}=0.049$ translates to an energy ratio of $\zeta_\mathrm{ep}=0.22$, which we adopt as the fiducial value for our \crest runs in order to compare it to the steady-state model without diffusion.
Since the CR physics included in the Fokker–Planck equation is independent of this normalisation, the resulting spectral shapes remain unaffected and can, in principle, be re-scaled to any desired injection efficiency in post-processing.

\section{Steady-state vs. CREST}
\label{Sec:Steady-state vs. CREST}

In this section, we compare the cell-based steady-state approach obtained with \crayon to the time-dependent electron spectra from \crest, which explicitly follow the temporal evolution of the CR electron population.

\subsection{Spatial distribution}\label{sec:SpatialDistribution}
\begin{figure}
    \centering
    \includegraphics[]{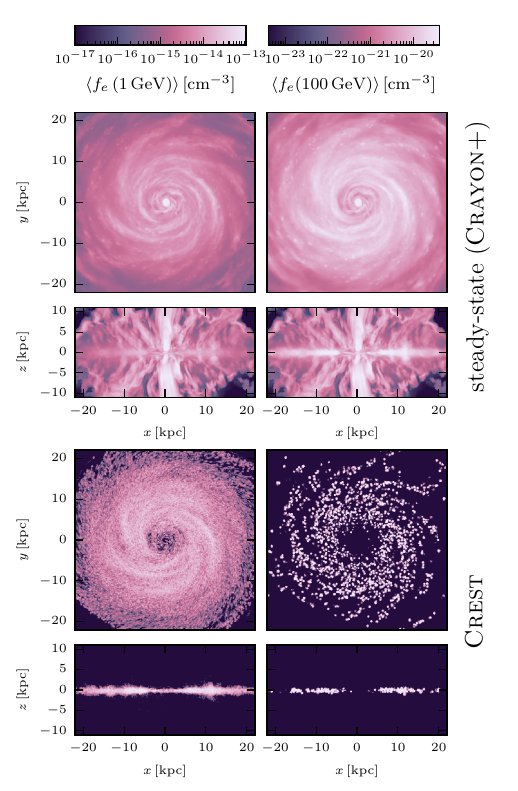}
    \caption{Distribution of CR electrons in the steady-state model with \crayon (upper four panels) and \crest (lower four panels), both face-on and edge-on at 1 and 100~GeV (left- and right-hand panels, respectively) at $t=3$~Gyr.}
    \label{fig:electron_maps}
\end{figure}

Figure~\ref{fig:electron_maps} shows maps of the volume-weighted mean CR electron spectrum $f(p)$ at $p=1~\mathrm{GeV}/(m_\mathrm{e}c^2)$ and $100~\mathrm{GeV}/(m_\mathrm{e}c^2)$ within a 4~kpc-thick slice, at $t=3$~Gyr, when the SFR is $4.4~\mathrm{M_\odot\,yr^{-1}}$.
The spatial distribution in the steady-state model is clearly more extended, both within the disk and vertically above and below it, than in the \crest model. This broader distribution arises because in the post-processed steady-state calculation, CR electrons are tied to CR protons via a fixed injected electron-to-proton ratio. Since CR protons in the simulation experience spatial diffusion, they extend well beyond the star-forming disk, imprinting a similarly broad distribution on the steady-state electron population.

In contrast, the CR electrons evolved with \crest are confined to a thinner disk, with the difference in morphology becoming increasingly pronounced at higher energies. This behavior is expected: as electron energy increases, the cooling time decreases strongly with energy, until radiative losses can no longer be balanced by continuous injection. High-energy electrons therefore deviate strongly from a steady-state configuration and are found only near their injection sites (i.e.\ recent SN explosions), where they cool quickly. We note, however, that including explicit spatial diffusion for CR electrons in \crest would potentially smooth out their distribution to some degree.

\begin{figure}
    \centering
    \includegraphics[]{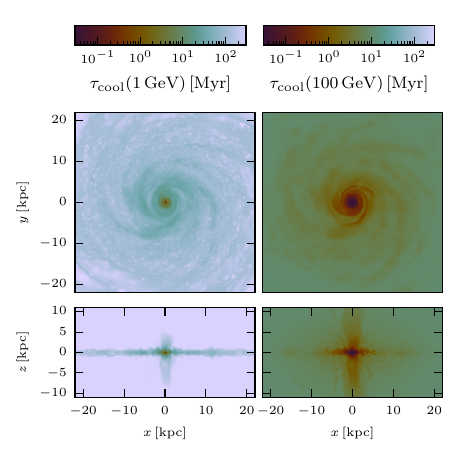}
    \caption{Slices showing cooling timescales for CR electrons at 1~GeV (left) and 100~GeV (right), shown both face-on (upper panels) and edge-on (lower panels) at $t=3$~Gyr. The timescales are calculated from slices of the gas density, magnetic field strength and photon energy density.}
    \label{fig:tau_cool_maps}
\end{figure}

To assess where spatial diffusion might become important, we compare cooling timescales to characteristic diffusion timescales from the literature.
Figure~\ref{fig:tau_cool_maps} shows maps of the total electron cooling time, $\tau_\mathrm{cool}=(\tau_\mathrm{sync}^{-1}+\tau_\mathrm{IC}^{-1}+\tau_\mathrm{Coul}^{-1}+\tau_\mathrm{brems}^{-1})^{-1}$, for CR electrons at 1~GeV and 100~GeV at $t=3$~Gyr.

Estimates for electron transport speeds vary widely.
\citet{2024Krumholz} infer a diffusion coefficient of $1.4^{+5.5}_{-0.8}\times10^{30}~\mathrm{cm^2\,s^{-1}}$ at TeV energies from gamma-ray emission around the globular cluster Terzan~5.
\citet{2025HopkinsXrayCGM} find an effective transport coefficient of $(0.8 - 2)\times10^{29}~\mathrm{cm^2\,s^{-1}}$ at GeV energies from X-ray halos,
while MW studies yield $\kappa_\parallel (E)\approx 2\times 10^{28}~\mathrm{cm^2\,s^{-1}}\times[E/(\mathrm{3~GeV})]^{\delta}$ with $\delta=0.5$ \citep{2010Strong, 2020Evoli}, corresponding to $\kappa_\parallel(1~\mathrm{TeV})\approx3.7\times 10^{29}~\mathrm{cm^2\,s^{-1}}$.
If we estimate the diffusion timescale from this last value via $\tau_\mathrm{diff}(E)=H^2/\kappa_\parallel(E)$, where $H$ denotes the CR gradient length, we infer $\tau_\mathrm{diff}(1~\mathrm{GeV})\approx650$~Myr for $H=5$~kpc and $100$~Myr for $H=2$~kpc, or at higher energies $\tau_\mathrm{diff}(100~\mathrm{GeV})\approx65$~Myr for $H=5$~kpc and $10$~Myr for $H=2$~kpc. If $\kappa_\parallel$ is an order of magnitude larger, the corresponding timescales decrease by a factor of ten.

Comparing these estimates to Fig.~\ref{fig:tau_cool_maps}, we find that within most of the galactic disk (with $R<10$~kpc and $z<1$~kpc), the cooling timescale at 1~GeV is $\lesssim 50$~Myr, shorter than typical diffusion times inferred for the MW. Thus, cooling dominates over spatial transport in the disk.
At larger heights and in the outer disk, however, $\tau_\mathrm{cool}$ can exceed 100~Myr, implying that diffusion may become important in those regions, especially if the larger effective coefficients suggested by \citet{2025HopkinsXrayCGM} apply.

At 100~GeV (shown in the right-hand panel of Fig.~\ref{fig:tau_cool_maps}), the cooling becomes far more efficient, due to the strong energy dependence of IC and synchrotron cooling. The cooling time is $\tau_\mathrm{cool}\lesssim1$~Myr in the central disk and $\lesssim10$~Myr even in the outskirts or above and below the disk. Hence, for such high-energy electrons, spatial diffusion is unlikely to play a significant role, as radiative losses occur much faster than transport.
It is also worth noting that observations and simulations indicate that CR diffusion can be strongly suppressed in the vicinity of sources, as inferred from TeV halos around pulsars \citep{2017Abeysekara} and SNRs \citep{Hanabata2014}, and from hybrid simulations showing self-generated regions of reduced diffusivity around CR sources \citep{2021Schroer, 2022Schroer}. This would further reduce the importance of spatial diffusion.

\subsection{Global electron spectrum}\label{sec:GlobalSpectra}

\begin{figure*}
    \centering
    \includegraphics[]{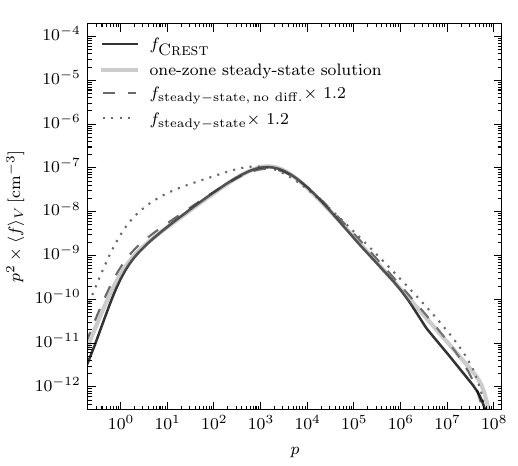}
    \includegraphics[]{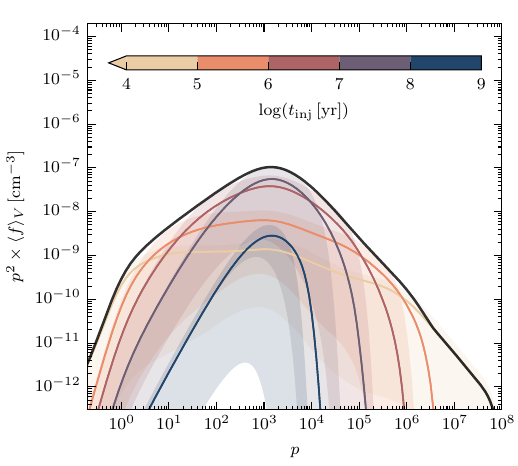}
    \caption{\textit{Left:} CR electron spectra in the galactic disk ($R<18$~kpc, $|z|<1$~kpc) at $t=1$~Gyr. The \crest spectrum (black) closely follows the one-zone steady-state model (light gray) up to $p\approx 10^4$, then steepens slightly, and deviates noticeably at $p\gtrsim 10^6$. The cell-based steady-state spectrum without diffusion (dashed), obtained with \crayon, matches the one-zone result, while diffusion (dotted) flattens the spectrum at low momenta ($p\lesssim 10^3$). 
    \textit{Right:} The \crest spectrum (black) decomposed by time since the last injection ($t_\mathrm{inj}$), as indicated by the colorbar. The individual components are volume averaged spectra, rescaled to the total volume. The total spectrum at $p=10^3$ is dominated by older electrons ($t_\mathrm{inj}\sim10^7-10^8$~yr), whereas the high-energy end is composed entirely of young electrons that have recently undergone energy injection ($t_\mathrm{inj}\lesssim10^5$~yr). Shaded regions show the 16th-84th percentile range of each electron population.}
    \label{fig:electron_spectra}
\end{figure*}

In Fig.~\ref{fig:electron_spectra}, we compare the total CR electron spectra within the galactic disk ($R<18\,\mathrm{kpc}$ and $\lvert z \rvert < 1\,\mathrm{kpc}$\footnote{This region encompasses 99.9\% of the total SFR.}) for different models at $t=1$~Gyr.
We analyse the simulation 500~Myr after we initialised the tracers (at $t=500$~Myr) as the CR electron spectrum has fully converged by this time (see Section~\ref{sec:Memory}); we further discuss its evolution up to $t=4$~Gyr in Section~\ref{sec:TemporalEvolution}.

We show $f_\mathrm{steady-state}$, i.e.\ the electron spectrum in the cell-based steady-state approach including both advection and diffusion, as well as $f_\mathrm{steady-state,no\ diff}$ which neglects diffusion. In the latter, we neglect diffusion for both protons and electrons. We discuss the effect of only neglecting diffusion for electrons while still accounting for diffusion of protons in App.~\ref{App:diffusion_e_p}. We note that neglecting both the advection and diffusion term in the cell-based steady-state approach leads to an identical total spectrum within the galactic disk in comparison to the spectrum where we only neglected diffusion. These models only differ at low momenta $p\lesssim 10^3$, where diffusion flattens the spectrum.

To facilitate comparison of spectral shapes, the steady-state spectra are scaled by a factor of 1.2 after adjusting for the fitted electron-to-proton ratios (see App.~\ref{App:diffusion_e_p} and Table~\ref{tab:kep}). 
This rescaling indicates that roughly 20\% more electron energy would be required in the cell-based steady-state normalisation in order to match the \crest results, implying that post-processing steady-state models seem to slightly overestimate the injected energy that is required to match the full time-dependent evolution of the spectra.
In addition, we show a one-zone steady-state spectrum (see Eq.~\ref{equ:f_steady}), which is normalised using the simulation's SFR of $14.5~\mathrm{M_\odot\,yr^{-1}}$, using Eq.~\eqref{eq:C_dot_tilde}\footnote{This yields a volume averaged injection rate of $\dot{\tilde{C}}_\mathrm{inj}=\dot{C}_\mathrm{inj}/V_\mathrm{disk}=4.18\times 10^{-22}\,\mathrm{s^{-1}\,cm^{-3}}$ with $V_\mathrm{disk}=2h_\mathrm{disk}\pi R_\mathrm{disk}^2$.}. 
It furthermore assumes mean disk properties of $B=3.7\mu\mathrm{G}$ and $\varepsilon_\mathrm{ph}=8\times \varepsilon_\mathrm{CMB}$. However, matching the spectral shape of the \crest spectrum requires 1.4 times the averaged gas density in the disk ($n_\mathrm{gas}=0.39\,\mathrm{cm^{-3}}$), suggesting that the global \crest spectrum effectively reflects a steady-state corresponding to denser environments, as we will discuss below.

Overall, we find that the global \crest spectrum closely resembles the shape of a steady-state spectrum across a large range in momenta. 
At low momenta $p\lesssim5\times10^{3}$, the \crest spectrum almost perfectly follows the one-zone steady-state model.
However, as mentioned above, this is only the case when we double the average gas density in the one-zone steady-state model. This adjustment is consistent with CR injection at SN sites in dense, star-forming regions. Consequently, low-energy electrons in \crest experience enhanced Coulomb cooling relative to the cell-based steady-state model. This behaviour arises, however, because the effective ISM model \citep{2003SpringelHernquist} does not explicitly capture the environment of SN explosions. Adopting a more detailed multiphase ISM treatment and higher mass resolution could alter the characteristic densities at SN injection sites, and thereby modify this conclusion \citep[see e.g. figure~5 of][]{2025Sike}.
At higher momenta, the total \crest spectrum steepens slightly in comparison to the steady-state models, before it steepens markedly above $p\sim2\times10^{6}$.
This is because at these large momenta, the spectrum becomes dominated by freshly injected electrons that have not yet reached equilibrium, as we will argue below.

The right-hand panel of Fig.~\ref{fig:electron_spectra} shows the contribution of tracer particles binned in different times since their last injection event, $t_\mathrm{inj}$, to the total electron spectrum obtained with \crest. This figure illustrates the build-up of a quasi-steady-state spectrum as the superposition of electrons with different ages. We find indeed that the total spectrum at $p\gtrsim 2\times 10^{6}$ is dominated by electrons with recent injection events, i.e.\ $t_\mathrm{inj}<10^{5}\,$yr (yellow line). 

\begin{figure}
    \centering
    \includegraphics[]{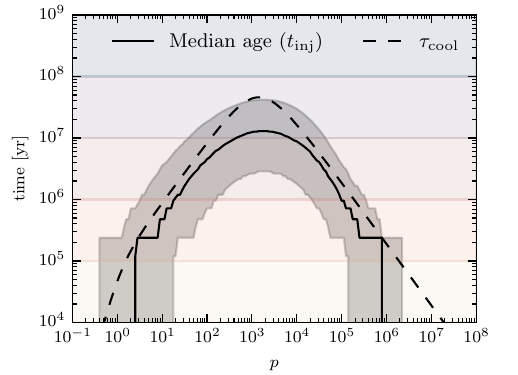}
    \caption{Median ages of CR electrons contributing to the total spectrum shown in Fig.~\ref{fig:electron_spectra} (solid lines) compared to their cooling timescales estimated from disk-averaged gas properties (dashed line). The shaded regions denote the 16th–84th percentile range and overlap closely with the cooling time up to $p\sim2\times10^6$. At higher momenta, the median age drops below the cooling time, indicating that these electrons were recently injected and have not yet reached a steady-state. The shaded horizontal bands correspond to the $t_\mathrm{inj}$ bins shown in Fig.~\ref{fig:electron_spectra}.}
    \label{fig:median_ages_tau_cool}
\end{figure}

The steady-state assumption requires that the typical timescale between electron injection events balances the energy-loss timescale from cooling.
We test this condition in Fig.~\ref{fig:median_ages_tau_cool} by comparing the spectral contribution-weighted median age of CR electrons to their cooling time, which includes bremsstrahlung, Coulomb, synchrotron and IC losses.

Across most of the momentum range, the median electron age tracks the cooling time closely, confirming that injection and cooling are approximately in equilibrium. 
However, at momenta above $p\gtrsim2\times10^6$, the median age falls sharply below the cooling time. This indicates that the highest-energy electrons are dominated by recently injected populations such that a steady-state between injection and cooling has not yet been established.
At these large momenta, the cooling times are extremely short, with $\tau_\mathrm{cool}<7\times10^4$~yr, based on disc-averaged gas and radiation energy densities. Consequently, the associated high-energy electrons cool before they can occupy large volumes. 
Because our model does not include explicit spatial diffusion, this confinement is physical, though it is likely enhanced by numerical effects related to limited mass and spatial resolution. 
In particular, both the spatial and temporal clustering of SN events artificially amplify this localization: with a gas mass resolution of 
$10^4\, \mathrm{M_\odot}$, and a SN rate of one per $100~\mathrm{M_\odot}$ of formed stars, a single star-formation event in the simulation represents roughly 100 SN explosions. As a result, CR injection occurs in bursts concentrated in both space and time, reducing the effective injection duty cycle and suppressing the large-scale filling factor of freshly accelerated electrons. 
This combination of rapid cooling and clustered injection naturally explains the observed steepening and reduced normalization of the total electron spectrum above $p\approx2\times10^6$.

The spectral peak at around $p\sim2\times10^3$ is instead dominated by older electrons, which experienced an injection event $10^7$ to $10^8$ years ago. This is because at those momenta, the cooling timescales are typically the longest, reaching cooling times of up to $\sim5\times10^7$~years in the disk.
The oldest electrons in the simulation with $t_\mathrm{inj}>10^8$~yr contribute little to the total spectrum, consistent with their ages being longer than the cooling times across all momenta (see Fig.~\ref{fig:median_ages_tau_cool}).
Finally, we note that the qualitative agreement between the global \crest and steady-state spectra persists across all analysed simulation times.

\subsection{Electron spectra in different regions}

\begin{figure*}
    \centering
    \includegraphics[]{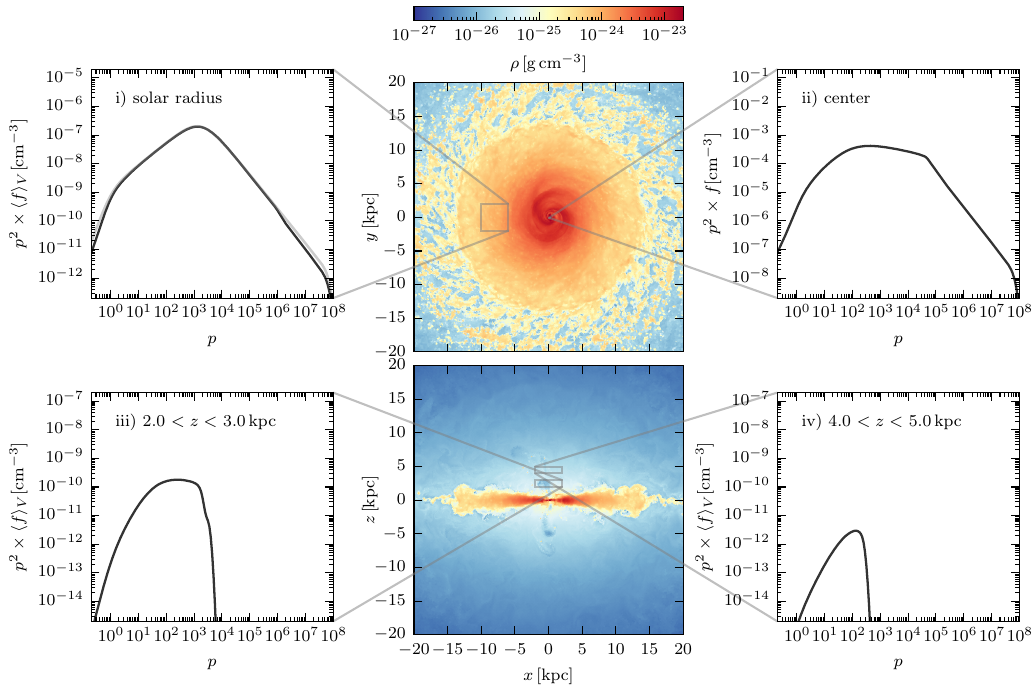}
    \caption{CR electron spectra in four representative regions of the galaxy at 0.9~Gyrs: Panel i) shows the averaged spectrum from \crest (black line) around the solar radius ($6~\mathrm{kpc}<R<10~\mathrm{kpc}$ and $\lvert z\rvert <1~\mathrm{kpc}$), which closely resembles a simple one-zone steady-state model (grey line). 
    In panel ii), the spectrum of a selected tracer from the central region is shown that experienced a recent injection event (located at $x=0.13$, $y=-0.05$, $z=0.02$~kpc). It shows both the injected slope ($p^{-2.2}$) and a developing steady-state tail ($p^{-3.2}$).
    In the outflow of the galaxy, as shown in panels iii) and iv), the averaged spectra within a cylinder of radius $R=2.5$~kpc show progressively cooled spectra with increasing height above the disk.
    The middle panels show corresponding gas density maps (face-on and edge-on). Besides the differences in shape we note strong variations in amplitude across the different regions.}
    \label{fig:map_spectra}
\end{figure*}

The global spectrum presented above shows that CR electrons in the disk approximately follow a steady-state distribution, with deviations only at high momenta. Hence, globally, the steady-state assumption seems to be a relatively good approximation. 
We now examine whether this behaviour also holds locally across the galaxy.
For this analysis, we select the snapshot at $t = 0.9$~Gyr, taken at an early stage of the CR-driven outflow. Similar qualitative features are found at later evolutionary stages.

Figure~\ref{fig:map_spectra} illustrates electron spectra in four representative regions of the simulated galaxy. 
First, panel (i) shows the spectrum averaged around the solar radius ($6<R<10$~kpc, $\lvert z\rvert<1$~kpc), where the spectrum closely resembles a one-zone steady-state solution, aside from a mild steepening above $p\gtrsim10^6$.
Panel (ii) presents a single tracer selected from the central region that experienced a recent SN injection. Its spectrum retains the injected power-law slope of $p^{-2.2}$ between $10^2\lesssim p\lesssim10^5$, transitioning toward the steady-state slope of $p^{-3.2}$ at higher momenta, capturing the formation of a steady-state spectrum in real time \citep[see also figure 6 of ][]{2019Winner}.
Panels (iii) and (iv) show electron spectra in the galactic outflows, 2–3~kpc and 4–5~kpc above the disk. Both exhibit strongly cooled, aged spectra, consistent with advection-dominated transport \citep[see also figure 4 of ][]{2019Winner}. The cutoff shifts from $p\sim5\times10^3$ at 2–3~kpc to $p\sim4\times10^2$ at 4–5~kpc, reflecting continued radiative cooling as tracers move away from the disk. 
Since CR injection is restricted to SN remnants within the star-forming disk, and no additional (re-)acceleration mechanisms such as resolved shocks (Fermi-I) or turbulence (Fermi-II) are included, these outflow spectra naturally exhibit strong cooling signatures.
The inclusion of such processes will be explored in future work.
Together, these examples spectra demonstrate that while the \textit{global} electron spectrum resembles a steady-state distribution, individual regions show pronounced deviations depending on the local injection history and environment.

\begin{figure*}
    \centering
    \includegraphics[]{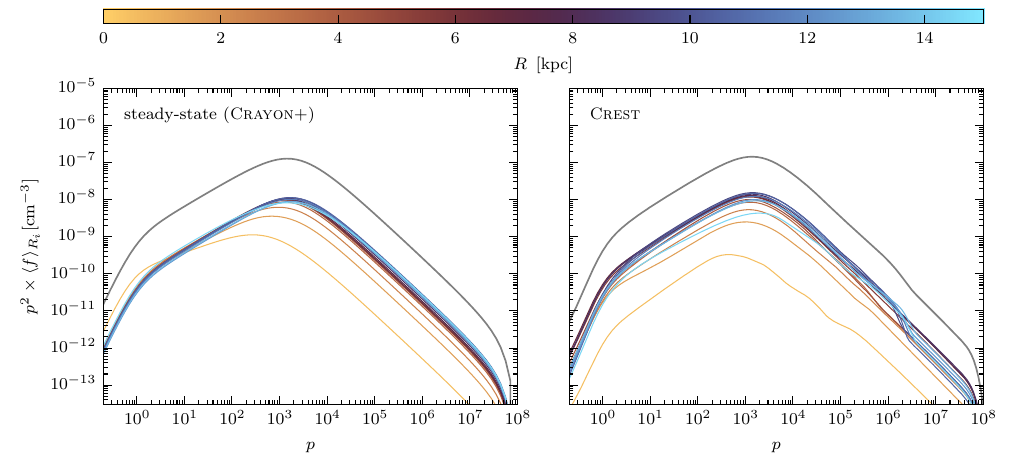}
    \caption{Radially binned CR electron spectra in the cell-based steady-state model with \crayon (left-hand panel) and in \crest (right-hand panel). The spectra are shown in 15 cylindrical bins within $\lvert z \rvert<$1~kpc. The total spectrum in this region (grey) is overplotted for reference.}
    \label{fig:spectra_bins}
\end{figure*}

To examine spatial variations more systematically, Fig.~\ref{fig:spectra_bins} compares radially binned spectra from the steady-state (\crayon) and time-dependent (\crest) models within a 2~kpc-thick disk.
In the steady-state case, spectral shapes are nearly identical across all radii, differing only in the peak position, particularly in the central regions, where Coulomb, IC, and synchrotron losses intersect at different momenta.
In contrast, the \crest\ spectra show mild regional variation, maintaining largely steady-state-like shapes but with small deviations, most notably a steepening at $p\gtrsim10^6$ as discussed in Section~\ref{sec:GlobalSpectra}. This demonstrates that averaging over large enough spatial regions in the \crest model recovers steady-state like spectra, with only small deviations in the spectral shapes.

\subsection{Temporal evolution of the electron spectrum}\label{sec:TemporalEvolution}

\begin{figure}
    \centering
    \includegraphics[]{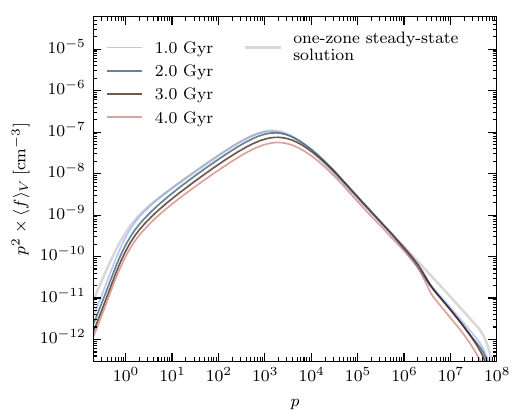}
    \caption{CR electron spectra from \crest at four different times (1--4~Gyr), averaged over the galactic disk ($R<18$~kpc, $\lvert z \rvert<1$~kpc).}
    \label{fig:spectra_times}
\end{figure}

Figure~\ref{fig:spectra_times} shows the temporal evolution of the disk-averaged CR electron spectrum between 1 and 4~Gyr. 
In all four cases, the simulation is restarted from a snapshot 500~Myr earlier with one tracer particle per cell. 
Given the short electron cooling times, this is sufficient to recover the full electron spectrum with high accuracy (as we show explicitly in Section~\ref{sec:Memory}).

Below $p\lesssim10^6$, the spectral shape remains remarkably stable across time, with only a slight decrease in normalisation at $p\lesssim10^4$. At larger momenta, the normalisation is nearly unchanged despite the substantial decrease in the SFR, i.e.\ from $14.5$ to $2.9~\mathrm{M_\odot\,yr^{-1}}$.

This behaviour arises because IC cooling dominates at high energies in most of the disk (see last column in Fig.~\ref{fig:map}), and the IC loss term scales with the local radiation field, which in turn tracks the SFR.
As both, the cooling and injection rates decline together, the resulting steady-state normalisation remains approximately constant over time at $p\gtrsim10^4$, where IC cooling dominates.
This effect is reproduced in Appendix~\ref{app:steady-state-exp-q} by a one-zone test model with exponentially declining injection and photon energy density (see Fig.~\ref{fig:steady-state-injRate}).

At low momenta, the normalisation decreases more slowly than the SFR: between 1 and 4~Gyr, the SFR drops by a factor of $\sim5$, while the low-energy normalisation decreases by only a factor of $\sim$2. 
This can be explained by a reduction in the mean gas density in the disk over time, leading to weaker Coulomb losses at low energies, thus partially compensating for the lower CR injection rate.

At the highest momenta $p\gtrsim10^6$, all spectra steepen relative to a simple one-zone steady-state model (gray line), albeit by a different amount. 
As shown in Fig.~\ref{fig:electron_spectra}, these high-energy electrons are dominated by young populations with recent injection events ($t_\mathrm{inj}\lesssim10^5$~yr).
These populations have not yet reached equilibrium, which explains both the temporal variability and the systematic steepening of the high-momentum tail.

\subsection{Comparison to observed electron data}

\begin{figure*}
    \centering
    \includegraphics[]{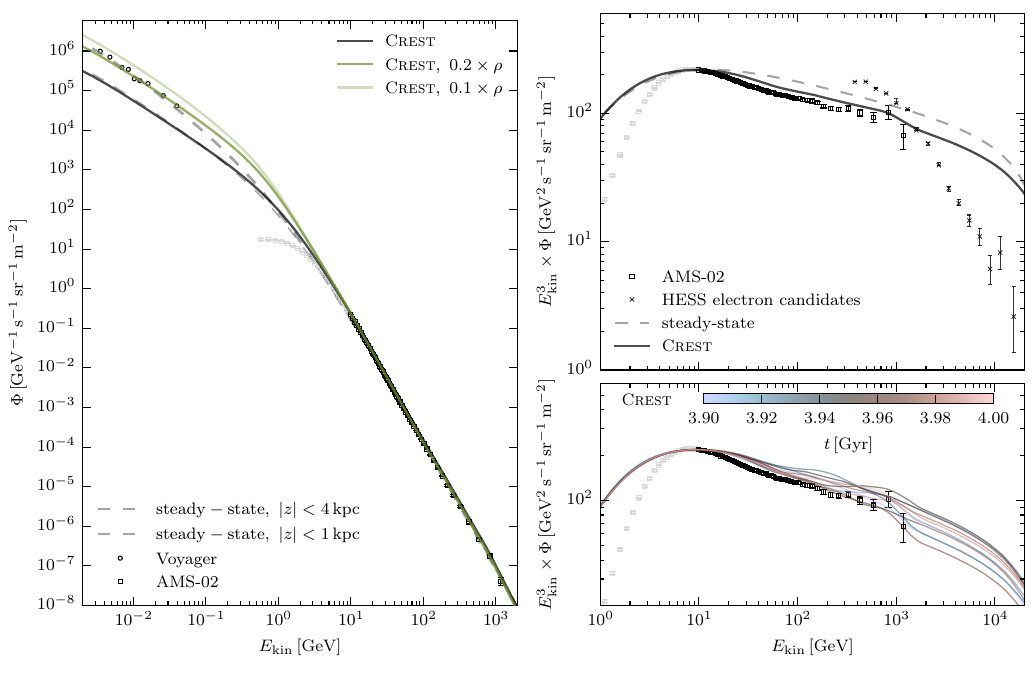}
    \caption{CR electron spectra from \crest (solid lines) and the steady-state \crayon model (dashed lines), compared to observations form Voyager~1 \citep{Cummings2016}, AMS-02 \citep{Aguilar2014b} and H.E.S.S. \citep{2024Aharonian}. 
    The H.E.S.S. data likely include residual contamination from CR nucleons and should thus be interpreted as an upper limit on the true electron flux. To facilitate a comparison of spectral shapes, the \crest and steady-state spectra are rescaled to the AMS-02 data at 10~GeV (see text for details).
    \textit{Left:} Comparison of simulated spectra around the solar radius ($6<R<10$~kpc) with observations. While AMS-02 data below 10~GeV (grey squares) are affected by solar modulation, Voyager~1 probes the unmodulated low-energy spectrum. The green (light green) line shows a \crest rerun with the gas density reduced by a factor of five (ten) to mimic the low-density environment of the Local Bubble, improving the match to Voyager~1.
    \textit{Upper right:} Spectra multiplied by $E_\mathrm{kin}^3$ to highlight differences in spectral slopes. The \crest spectrum (black) is slightly steeper than the steady-state result and aligns more closely with AMS-02 at $E_\mathrm{kin}\gtrsim10$~GeV, though both models overpredict the flux at $\gtrsim1$~TeV relative to H.E.S.S.
    \textit{Lower right:} Temporal variability of the \crest spectra over ten consecutive snapshots from 3.9–4.0~Gyr. Above $\sim$100~GeV, intermittent injection and rapid cooling lead to noticeable changes in the spectral shape.
    }
    \label{fig:spectra_AMS}
\end{figure*}

To test the consistency of our simulated CR electron spectra with observations, we compare the resulting spectra to direct measurements from local observations.
Figure~\ref{fig:spectra_AMS} compares the simulated CR electron spectra around the solar radius ($6~\mathrm{kpc}<R<10$~kpc) with observational data from Voyager~1 \citep{Cummings2016}, AMS-02 \citep{Aguilar2014b} and H.E.S.S, an imaging atmospheric Cherenkov telescope \citep{2024Aharonian}. The H.E.S.S. fluxes may still contain background contamination from CR nucleons and therefore represent an upper limit on the true electron flux.
We analyse the final simulation snapshot at $t=4$~Gyr, when the SFR has declined to 2.9~$\mathrm{M_\odot\,yr^{-1}}$.
To facilitate comparison of spectral shapes, we renormalise the simulated spectra at $E_\mathrm{kin}=10$~GeV to match the AMS-02 data by applying a scaling factor of 0.1. 

This offset is partly attributable to the simulation’s higher SFR compared to the MW, whose current SFR is estimated between $1$–$2~\mathrm{M_\odot\,yr^{-1}}$ \citep[e.g.][]{2011Chomiuk, Licquia_2015}, and potentially as low as $0.67~\mathrm{M_\odot\,yr^{-1}}$ \citep{2025Quintana}. Accounting for this range implies a rescaling factor between 0.23 and 0.68 purely from the SFR difference. 
Moreover, this range lies within the uncertainties in both the number and characteristic energy of SNe per unit stellar mass formed, which are not well constrained.
Any residual discrepancy likely arises from local environmental effects: while the Voyager measurements probe the low-density Local Bubble, our simulated spectra represent averages over a several-kpc-wide slab of the disk. A simulation with both a much higher resolution and an improved ISM model would be required to study this in more detail.

At energies above 10~GeV, both the \crest and \crayon (steady-state) spectra agree well with AMS-02 observations (left-hand panel of Fig.~\ref{fig:spectra_AMS}).
At lower energies, however, only the steady-state model reproduces the Voyager~1 data when averaged over a vertical extent of $\pm4$~kpc around the disk.
Reducing the integration height in the steady-state model (for example to $\pm1$~kpc, as shown in Fig.~\ref{fig:spectra_AMS}) leads to an underprediction of the flux below $\sim$1~GeV. This is due to the strong dependence of Coulomb cooling on gas density, which decreases rapidly with height. 

In contrast, we find that the \crest spectra show little variation with height (we display the $\lvert z \vert <1$~kpc case in Fig.~\ref{fig:spectra_AMS}). 
This insensitivity likely stems from the fact that \crest injects energy into electrons in dense, star-forming regions where Coulomb-losses are efficient, causing low-energy electrons to cool before escaping to lower-density regions (at average gas densities in the disk, the Coulomb cooling time is $<10^6$ years for $p<20$ or $E_\mathrm{kin}<10^{-2}$~GeV; see Fig.~\ref{fig:median_ages_tau_cool}). Consequently, Coulomb cooling may be somewhat overestimated in the current \crest implementation. 
Furthermore, this effect is compounded by the limited spatial resolution and the use of a pressurized ISM model \citep{2003SpringelHernquist}, which does not resolve the multiphase ISM structure. Improving on this will probably moderate the strength of Coulomb losses along electron trajectories. 

Additionally, the absence of spatial diffusion in \crest further contributes to this discrepancy. As shown in Fig.~\ref{fig:electron_spectra}, including diffusion in the steady-state model primarily affects momenta $p<10^3$, corresponding to kinetic energy below $\sim$0.5~GeV, i.e.\ the energy range probed by Voyager~1. 
Finally, environmental differences may once again play a role. The Local Bubble, where Voyager measures the spectrum, is underdense in comparison to the average ISM. The observationally inferred electron density of 0.012~$\mathrm{cm^{-3}}$ \citep{2021Linsky} in the Local Bubble is well below the Galactic average of $\sim$1~$\mathrm{cm^{-3}}$.
In our simulation, the mean hydrogen number density at the solar radius is $\approx0.16~\mathrm{cm^{-3}}$.

To illustrate this effect, we rerun \crest using the same tracer data but artificially reduce the gas density to mimic the low-density environment of the Local Bubble. The resulting spectra (green lines in Fig.~\ref{fig:spectra_AMS}) show improved agreement with the Voyager~1 data, supporting the interpretation that environmental density differences largely explain the remaining mismatch at low energies.
The fact that we can not match the Voyager data with average ISM gas densities, but require a lower density, could imply that the electrons at low energies originate from inside the Local Bubble. If this is the case, that is if the Voyager spectrum indeed reflects the conditions within the Local Bubble, this could potentially provide a lower limit on its age: the bubble must be older than the cooling time of low-energy electrons, otherwise the local, low-density environment would not yet have affected its spectral shape. An electron density of $0.012~\mathrm{cm^{-3}}$ would imply a Coulomb cooling timescale of $\approx10$~Myr at $E_\mathrm{kin}=2\times10^{-3}$~GeV, which is consistent with age estimates by \citet{2022Zucker}.

To better highlight differences in the spectral slope at high energies, the right-hand panels of Fig.~\ref{fig:spectra_AMS} show the measured AMS-02 flux multiplied by $E_\mathrm{kin}^3$.
The upper panel compares the time-averaged \crest spectrum (averaged over ten snapshots from $3.9$ to $4.0$~Gyr) with the steady-state \crayon model, which is time-independent over this interval.
The \crest\ spectrum exhibits a mild steepening above $\sim$10~GeV relative to the steady-state result, bringing it closer to the AMS-02 data, although some discrepancies remain. 
At the highest energies ($E_\mathrm{kin}\gtrsim1$~TeV), all models overpredict the flux compared to H.E.S.S., whose data show a strong steepening with energy, which is not captured by either model.
This overprediction may partly arise from our simplified injection prescription, which assumes a single power-law electron spectrum up to a sharp cut-off at 20~TeV (see Eqs.~\ref{eq:q(p)} and~\ref{eq:tilde_q(p)}). In reality, the spectrum of electrons escaping from SNRs is likely more complex. For instance, it might exhibit a broken power-law shape that steepens relative to the proton spectrum due to radiative losses or time-dependent acceleration efficiency \citep[e.g.\ ][]{2019DiesingCaprioli, 2021Cristofari, 2021MorlinoCelli}. We will explore the impact of such injection spectra, including broken power-law shapes, in future work.

The lower panel displays the time variability among the ten \crest snapshots. Above $\sim$100~GeV, the spectra vary significantly with time due to intermittent energy injection events and fast cooling at these energies, producing subtle changes in the spectral shapes. 
One snapshot ($t=3.97$~Gyr) matches the AMS-02 data particularly well at the highest energies. However, some fine spectral features are still not reproduced.
For instance, \citet{2020bEvoli} attributed the observed hardening of the CR electron spectrum at energies $\gtrsim$42~GeV to the transition into the Klein-Nishina regime of IC scattering on ultraviolet photons, which is not included in our current radiation field model.

Despite these limitations, the \crest spectra successfully reproduce the overall spectral shape and its evolution up to TeV energies, capturing the main physical mechanisms that shape the observed CR electron spectrum in the MW.

\section{The short memory of CR electrons}\label{sec:Memory}

\begin{figure*}
    \centering
    \includegraphics[]{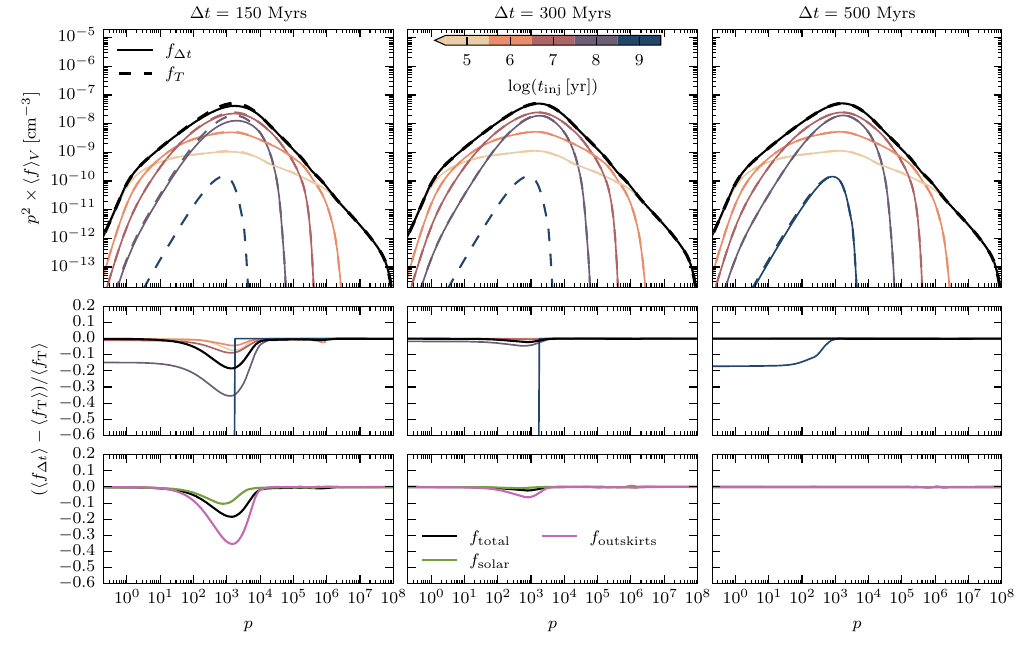}
    \caption{CR electron spectra at $t=1.2$~Gyrs for the full medium-resolution simulation (1M; solid lines), $f_T$, compared to restarted \crest runs in which all spectra were reset to zero at a time $\Delta t$ before the analysis time (dashed lines). 
    The upper panels are colour-coded by the time of the last injection event in each tracer particle, $t_\mathrm{inj}$, as indicated by the colour bar. 
    Spectra are averaged within a disk with radius $R=18$~kpc and $h\pm2$~kpc. 
    The second and third rows show the relative error introduced by restarting, binned by $t_\mathrm{inj}$ (second row) and averaged over different radial regions, i.e.\ around the solar radius ($7<R<9$~kpc; green) and in the outskirts ($12<R<15$~kpc; pink).}
    \label{fig:memory}
\end{figure*}

\begin{figure*}
    \centering
    \includegraphics[]{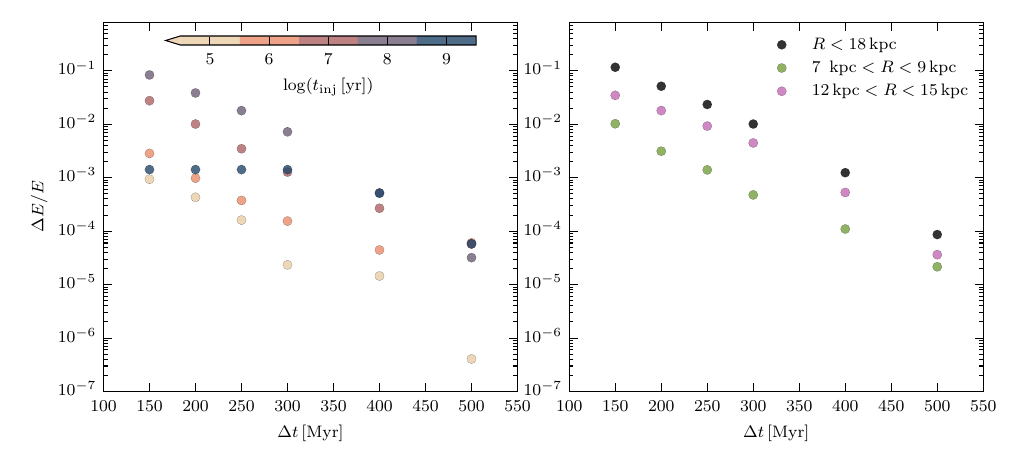}
    \caption{Relative error in CR electron energy for the same regions as in Fig.~\ref{fig:memory}.}
    \label{fig:memory_energy}
\end{figure*}

In this section, we quantify the `memory' of CR electrons. That is, we quantify the time span over which past injection and cooling must be tracked to accurately model the present-day spectra. 
Because electron cooling times vary strongly with energy and with local gas and magnetic field conditions within the galactic disk, it is not obvious a priori how far back in time the relevant history extends, in order to accurately model the full electron spectrum. 

To this end, we perform a full simulation to $T=1.2$~Gyr, evolving tracer particles from 0.2~Gyr onward.
To assess the required integration time for accurate CR electron spectra, we conduct additional runs that are restarted at earlier times $T-\Delta t$, with $\Delta t=\{150,\,300,\,500\}$~Myr. In each restart, tracer particles are placed in all gas cells within $R<50$~kpc of the galactic centre and evolved to $T$, recording all quantities required by \crest in post-processing.
To allow for multiple realisations, these tests are performed at medium resolution (`1M)', for which the global CR electron spectra are fully converged (see Fig.~\ref{fig:CRspectrum_resolutions}).

Figure~\ref{fig:memory} compares the electron spectra from the full run ($f_T$) to those from the restarted runs ($f_{\Delta t}$).
The total spectra in all of the restarted simulations closely resemble the total spectrum from the full simulation. 
Already after a restart time of $\Delta t=150$~Myr, the relative difference (shown in the second row) of the total spectrum is $\leq18\%$ across all momenta\footnote{For numerical stability, the relative differences are only computed where $p^2\langle f\rangle>10^{-30}\,\mathrm{[cm^{-3}]}$ and the spectral slope is $>-5$.}. 
The only visible discrepancy occurs around $p\sim10^3$, reflecting the absence of electrons with their last injection event $t_\mathrm{inj}>150$~Myr in the restarted run.
Spectra from tracers with more recent injection events ($t_\mathrm{inj}<150$~Myr) are essentially identical to those in the full simulation.

For a longer restart interval of $\Delta t=300$~Myr, the total CR electron spectrum at the final time of 1.2~Gyr is fully recovered, with a maximum deviation across all momentum bins of only $2.2\%$. 
Electrons that are injected earlier than 300~Myrs before the analysis time contribute negligibly to the total spectrum in the disk.

Finally, restarting from $\Delta t=500$~Myr fully recovers the electron spectrum from the full run, with relative errors below $0.2\%$. 
Only a small contribution of the oldest electrons with an injection event more than half a Gyr ago are lacking.

The lowest panels of Fig.~\ref{fig:memory} quantify these deviations for specific spatial regions, i.e.\ around the solar radius (in green) and in the outskirts of the disk (in pink). 
Averaging over spatial regions reduces the error introduced by neglecting older electrons (i.e.\ older than the runtime of the simulation). This is because electrons with $t_\mathrm{inj}>10^{8}$~yrs do not significantly contribute to the averaged electron spectrum in any region.
Around the solar radius, the spectra converge to within $<1\%$ already for $\Delta t=300$~Myr, whereas in the outer disk, deviations up to $6\%$ remain due to a larger contribution of old electrons with longer cooling times. 
In both regions, restarting 500~Myr before the analysis time yields fully converged results.

To further quantify this short `memory' of CR electrons, we examine the total CR electron energy. Figure~\ref{fig:memory_energy} shows the relative error in the integrated energy, $\Delta E_i /E= [E_i(f_{\Delta T}) - E_i(f_{\Delta t})]/ E(f_{\Delta T})$, computed in bins $i$ of the time since the last injection event, $t_\mathrm{inj}$. 
For bins where $t_\mathrm{inj}>\Delta t$, the restarted simulations contain no electrons ($E_i(f_{\Delta t})=0$), so $\Delta E_i/E$ directly represents the contribution of electrons of that age in the full run.

In all cases, the energy error remains below $10^{-1}$ and drops below $<10^{-2}$ for $\Delta t>300$~Myr, in all bins of $t_\mathrm{inj}$.
The same holds for the relative error in total energy in the galactic disk, as shown in the right-hand panel of Fig.~\ref{fig:memory_energy}. For $\Delta t>200$~Myr ($\Delta t>300$~Myr), the relative error falls below $10^{-1}$ ($10^{-2}$) in the galactic disk. 
At the solar radius, errors are an order of magnitude smaller, whereas in the outskirts, they are larger by a factor of $\sim$2--3, consistent with longer cooling times in that region. 

\begin{figure*}
    \centering
     \includegraphics[]{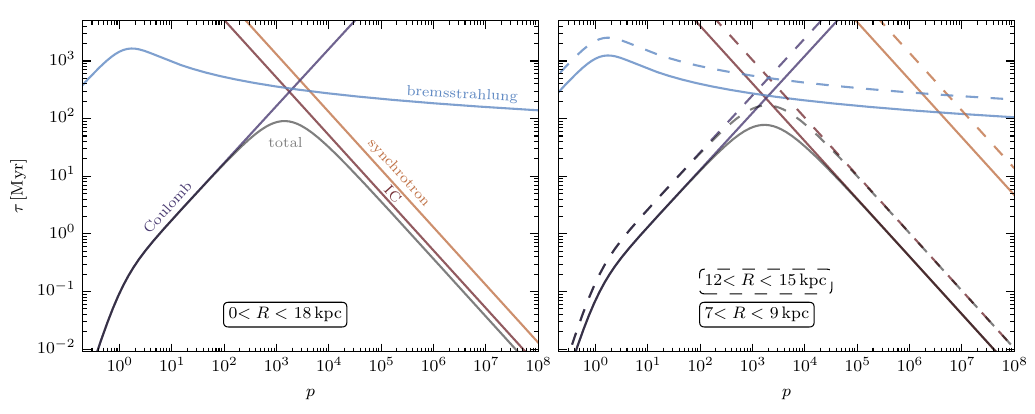}
    \caption{Cooling timescales as a function of electron momentum for Coulomb (purple), bremsstrahlung (blue), synchrotron (orange) and IC losses (red). The total cooling time (grey) combines all processes.
    Left: volume-averaged timescales across the galactic disk ($R<18$~kpc, $|z|<2$~kpc). The total cooling time peaks at $p\sim10^3$ with $\tau_\mathrm{cool}\approx100$~Myr and declines toward both smaller and larger momenta, where Coulomb and IC losses dominate, respectively.
    Right: timescales averaged over different radial regions. In the outer disk ($12<R<15$~kpc; dashed lines), weaker magnetic fields lead to weaker synchrotron losses, yielding total cooling times up to $\sim$170~Myr. Near the solar radius ($7<R<9$~kpc; solid lines), higher densities, field strengths, and photon energy densities shorten all loss processes by roughly a factor of two. This explains why the energy convergence in Fig.~\ref{fig:memory_energy} is reached twice as fast in the inner disk as in the outskirts.}
    \label{fig:timescales}
\end{figure*}

Figure~\ref{fig:timescales} illustrates the underlying cooling timescales as a function of momentum for the same simulation at $t=1.2$~Gyr.
The total cooling time $\tau_\mathrm{total}=[\tau_\mathrm{Coulomb}^{-1}(n_\mathrm{gas})+\tau_\mathrm{sync}^{-1}(B)+\tau_\mathrm{IC}^{-1}(\varepsilon_\mathrm{ph})+\tau_\mathrm{brems}^{-1}(n_\mathrm{gas})]^{-1}$ is computed from the volume-weighted averages of the relevant local quantities.

The left-hand panel of Fig.~\ref{fig:timescales} shows the averaged timescales in the galactic disk with a cylindrical radius of $R=18$~kpc and $\lvert{z}\rvert<2$~kpc. The total cooling time peaks at 90~Myr around $p\sim10^3$, corresponding to the peak of the total, disk-averaged electron spectrum in Fig.~\ref{fig:memory}. We note that we refer to a peak in $p^2f(p)$, which is the spectral contribution to the CRe energy density per logarithmic momentum interval in the relativistic regime.
At lower momenta, the cooling time declines steeply due to Coulomb losses, while at large momenta IC-losses dominate and strongly shorten the cooling time.

The right-hand panel of Fig.~\ref{fig:timescales} shows the same cooling timescales as a function of momentum but averaged in different spatial regions. Both are shown at large cylindrical radii, where the magnetic field strength has dropped significantly such that synchrotron cooling is strongly suppressed and much longer than IC cooling. 
In the outskirts of the galactic disk, with $12~\mathrm{kpc}<R<15$~kpc, the total cooling timescale peaks at around 170~Myr. At the solar radius, $7~\mathrm{kpc}<R<9$~kpc, the total cooling time is roughly by a factor of two shorter, consistent with denser gas, stronger magnetic field strengths, and higher photon energy densities, associated with a higher local SFR in this region.
This directly explains why the spectral and energetic convergence times in Figs.~\ref{fig:memory} and \ref{fig:memory_energy} are about twice as long in the outskirts compared to the solar neighbourhood.

In summary, CR electron populations in our simulated disk effectively lose memory of their injection and cooling history after $\sim$300~Myr.
Older electrons contribute negligibly to the present-day spectra or energy content, except in the low-density outskirts, where cooling times are longer.
This demonstrates that accurate time-dependent modelling of CR electrons in disk galaxies requires tracking their evolution over only a few hundred Myr, which is much shorter than typical galactic evolutionary timescales.

\section{Conclusion}
\label{Sec:Conclusion}

In this work, we have presented a study of the spectral evolution of CR electrons in MHD simulations of a MW–mass galaxy, using the moving-mesh code \arepo in combination with two complementary spectral models for the CR electron population. The time-dependent \crest code follows the full evolution of CR electron spectra on Lagrangian tracer particles, which record the MHD conditions and energy injection history along their trajectories. This approach enables us to solve the full time-dependent CR transport equation in momentum space, accounting for adiabatic effects as well as cooling processes, including synchrotron, IC, bremsstrahlung, and Coulomb losses.
We compared these time-dependent results to cell-based steady-state solutions with the \crayon code, which assumes a balance between injection and energy losses in each computational cell, as well as to simplified one-zone analytical steady-state models. This combination of methods allowed us to systematically assess when and where the steady-state assumption is valid and to identify regimes in which a time-dependent description becomes necessary.

We find that the global CR electron spectrum in the simulated galaxy remains close to a steady-state shape across most of the disk, with deviations only at high energies ($E \gtrsim 500~\mathrm{GeV}$, or $p \gtrsim 10^6$), where rapid synchrotron and IC cooling steepen the spectrum (see Fig.~\ref{fig:electron_spectra}). The median electron ages closely follow their cooling timescales up to this energy (Fig.~\ref{fig:median_ages_tau_cool}), confirming a near-equilibrium between injection and losses. 
Locally, however, \crest reveals strong spatial variations: recently injected electrons in star-forming regions exhibit non-equilibrium spectra, while outflow regions show strongly cooled, aged populations (Fig.~\ref{fig:map_spectra}). 
Comparing to observed data from Voyager~1, AMS-02, and H.E.S.S.\ (Fig.~\ref{fig:spectra_AMS}), both models reproduce the overall spectral shape above 10~GeV, with \crest providing a better match to the observed steepening at higher energies.  

By restarting simulations from different epochs (Section~\ref{sec:Memory}), we demonstrated that CR electrons retain only a short effective memory of their injection history, losing most of it within $\lesssim300$~Myr, consistent with their energy-loss timescales (Fig.~\ref{fig:timescales}). This implies that the observable CR electron population primarily traces the recent star-formation history of the galaxy. Our results therefore justify the use of steady-state models such as \crayon for global analyses, while highlighting the importance of time-dependent treatments with \crest to capture the spectral and spatial variations of high-energy CR electrons or galactic outflows.  

In future work, we will extend our analysis to compute the resulting non-thermal emission from synchrotron, IC, and bremsstrahlung processes, based on the live electron modelling with \crest. Comparing these predictions to those obtained from steady-state approaches will allow us to assess the validity and limitations of the steady-state assumption employed in \crayon, and to quantify its impact on observable radio and gamma-ray properties. 
While the simulations presented here are isolated, idealised disk galaxies, testing our approach in fully cosmological zoom-in simulations will be essential to verify the robustness of our results and to determine whether they hold in a more realistic galactic environment, such as in cosmological zoom galaxies incorporating CRs \citep[as e.g.\ in ][]{2021aHopkins,2024RodriguezMontero,2025Bieri}. In addition, we aim to extend our study across a broader range of galaxy masses to examine whether the similarities and differences between steady-state and live electron modelling found here for MW–mass systems persist in lower-mass galaxies.
Ultimately, this will inform whether steady-state modelling can be reliably applied to cosmological zoom simulations, or if the live electron modelling of \crest must be incorporated directly to capture the relevant physics in more realistic environments.
Moreover, we plan to explore simulations with more sophisticated, multi-phase ISM models and to incorporate two-moment CR transport schemes, such as the \textsc{Crisp} model \citep{2025ThomasPfrommerPakmor}, to achieve a more self-consistent treatment of CR propagation.
Finally, including Fermi-I acceleration and re-acceleration at resolved shocks in future simulations will allow us to investigate whether these processes can enhance the population of high-energy electrons in galactic outflows, potentially leading to distinct observational signatures in the radio and gamma-ray regimes.

\section*{Acknowledgements}
CP and JW acknowledge support from the European Research Council via the ERC Advanced Grant ``PICOGAL'' (project ID 101019746). JW acknowledges support by the German Science Foundation (DFG) under grant 444932369. RB is supported by the SNSF through the Ambizione Grant PZ00P2\_223532. PG acknowledges financial support from the European Research Council via the ERC Synergy Grant ``ECOGAL'' (project ID 855130). LJ acknowledges support from the Deutsche Forschungsgemeinschaft (DFG, German Research Foundation) as part of the DFG Research Unit FOR5195 – project number 443220636.

\section*{Data Availability}
The data underlying this article will be shared on reasonable request to the corresponding author.



\bibliographystyle{mnras}
\bibliography{literature}



\appendix

\section{Variations in the cell-based steady-state model}\label{App:diffusion_e_p}

\begin{figure}
    \centering
    \includegraphics[]{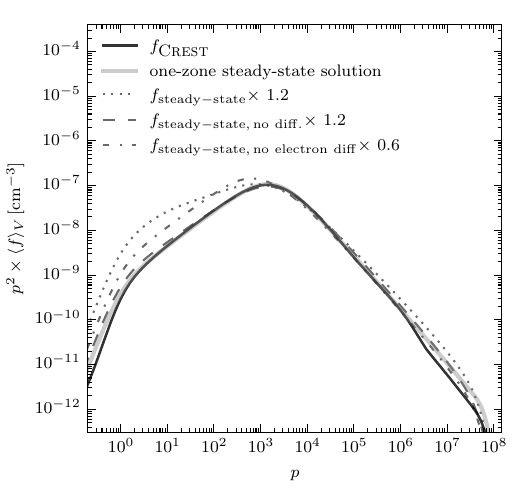}
    \caption{Different versions of the cell-based steady-state model with \crayon. This is the same as the left-hand panel of Fig.~\ref{fig:electron_spectra}, but additionally shows the case of excluding diffusion for CR electrons only (`steady-state, no electron diff’). The steady-state spectra have been renormalised so that we may better compare the spectral shapes (see text).}
    \label{fig:f_e_stst_nodiffel}
\end{figure}

We examine the impact of excluding the diffusion term in the steady-state equation (Equation~\ref{eq:diff-loss-equ}) solved in each computational cell by \crayon. In the fiducial model, the escape timescale combines advection and diffusion, $\tau_\mathrm{esc}^{-1}=\tau_\mathrm{diff}^{-1}+\tau_\mathrm{adv}^{-1}$, for both CR protons and electrons.
To isolate the role of spatial diffusion, we run two additional variants: one without diffusion for CR electrons only (`steady-state, no electron diff’), and one without diffusion for either species (`steady-state, no diff’). The resulting steady-state electron spectra are shown in Fig.~\ref{fig:f_e_stst_nodiffel}. As discussed in Section~\ref{sec:SpatialDistribution}, including spatial diffusion for CR electrons would likely smooth out their spatial distribution to some extend. In terms of the spectral shape, as discussed in Section~\ref{sec:GlobalSpectra}, including diffusion for CR electrons primarily affects the global spectrum at low momenta ($1 \lesssim p \lesssim 10^3$). When diffusion is excluded, Coulomb cooling dominates this regime, resulting in the spectra shown by the dashed and dash-dotted lines.
Neglecting diffusion for CR electrons alone leaves the overall spectral shape largely unchanged, but leads to a shift of the spectral peak in $p^2f(p)$ toward lower momenta and an increase in the overall normalisation (by roughly a factor of two) when diffusion of CR protons is retained in the post-processing calculation of the steady-state spectra (while the simulations themselves always include CR proton diffusion). This arises because the model includes energy-dependent diffusion: in high-density regions, proton spectra remain unaffected as hadronic losses dominate and are nearly energy-independent, while in low-density regions diffusion steepens the proton spectra where hadronic losses are weak. Since the spectra are renormalised to the local CR energy density after solving the steady-state equation, this steepening enhances the normalisation in low-density cells. Consequently, the primary CR electron spectra in those regions, where Coulomb cooling is weaker, also become more strongly normalised, shifting their peaks to lower momenta.
Those differences also lead to modest variations in the electron-to-proton ratio required to reproduce the observed Milky Way spectra. Table~\ref{tab:kep} summarises the corresponding values of $K_\mathrm{ep}^\mathrm{inj}$ for the different model variations, both when including secondary electrons and when considering only primary electrons. The inclusion or omission of diffusion changes the required ratio by roughly a factor of two, whereas accounting for secondary electrons has only a very minor effect. 

\begin{table*} 
\centering
\begin{tabular}{ l c c c c c }
\hline
Name                            & CRp diff & CRe diff & advection & $K_{\rm{ep}}^\mathrm{inj}$ & $K_{\rm{ep}}^\mathrm{inj}$ \\
                                &          &          &           & (incl. secondaries) & (only primaries) \\
\hline 
\hline
steady-state                   & \cmark\ & \cmark\ &  \cmark\ & 0.020 & 0.021 \\ 
steady-state, no diff          & \xmark\ & \xmark\ &  \cmark\ & 0.047 & 0.049 \\ 
steady-state, no electron diff      & \cmark\ & \xmark\ &  \cmark\ & 0.014 & 0.014\\
\hline
\end{tabular}
\caption{Injected electron-to-proton ratios for different variations of the cell-based steady-state model with \crayon, which are required to match the observed value in the MW around the solar radius.}
\label{tab:kep}
\end{table*}

\section{Steady-state tests}
\label{app:steady-state-exp-q}

To understand the shape of a steady-state spectrum with a non-constant injection term, we run a one-zone test model with a single tracer particle in \crest. 
We compare the result to the analytical steady-state solution resulting from a constant source function in time.

\subsection{Steady-state solution for a constant source function}

First, the solution for the spectrum in a steady-state between cooling and injection (without any advection or diffusion losses) is given by
\begin{align}
    f_\mathrm{steady}(p) = \frac{1}{\lvert \dot{p}(p) \rvert} \int_p^\infty q(p) \mathrm{d}p.
\end{align}
If the source function is constant in time and is given by
\begin{align}
q(p) = \frac{\mathrm{d}N}{\mathrm{d}p\, \mathrm{d}t\, \mathrm{d}V} =  \dot{\tilde{C}}_\mathrm{inj} p^{-\alpha_\mathrm{inj}} \theta(p-p_\mathrm{min}),
\end{align}
the steady-state solution simply becomes
\begin{align}
    f_\mathrm{steady}(p) = \frac{1}{\lvert \dot{p}(p) \rvert} \frac{\dot{\tilde{C}}_\mathrm{inj}}{\alpha_\mathrm{inj}-1} \times \begin{cases}
p^{1-\alpha_\mathrm{inj}} & p\geq p_\mathrm{min}\\
p_\mathrm{min}^{1-\alpha_\mathrm{inj}}& p < p_\mathrm{min}\\
\end{cases}.
\label{equ:f_steady}
\end{align}
We furthermore define the volume-integrated source function as 
\begin{align}
    Q(p) = \int q(p)\mathrm{d}V = \dot{C}_\mathrm{inj} p^{-\alpha_\mathrm{inj}} \theta(p-p_\mathrm{min}),
\end{align}
with $\dot{C}_\mathrm{inj}=\int \dot{\tilde{C}}_\mathrm{inj}\mathrm{d}V$.
The total luminosity injected into electrons is then given by
\begin{align}
    L_\mathrm{e}= \int_0^{\infty}Q(p)E_\mathrm{kin}(p) \dd p. 
\end{align}
Inserting the constant source function, this gives
\begin{align}
    L_\mathrm{e} = \dot{C}_\mathrm{inj}\int_0^{\infty} p^{-\alpha_\mathrm{inj}} \theta(p-p_\mathrm{min}) E_\mathrm{kin}(p) \dd p.
\end{align}
The integral can be solved with the incomplete beta function $\B_y(a,b)$ \citep{mabramowitz64:handbook} via
\begin{align}
    L_\mathrm{e}&=\int_0^{\infty}Q(p)\,E_\mathrm{kin}(p)\dd p=\frac{\dot{C}_\mathrm{inj}\,m_{\e}c^2}{\alpha_\mathrm{inj}-1}\\
&\quad\times\left[\frac{1}{2}\,\B_\frac{1}{1+p_\mathrm{min}^2}\left(\frac{\alpha_\mathrm{inj}-2}{2},\frac{3-\alpha_\mathrm{inj}}{2}\right)+
  p_\mathrm{min}^{1-\alpha_\mathrm{inj}}\left(\sqrt{1+p_\mathrm{min}^2}-1\right)\right]\nonumber\\
&\equiv \dot{C}_\mathrm{inj}\,m_{\e}c^2 A_\mathrm{bol} (p_\mathrm{min},\alpha_\mathrm{inj}).
\end{align}

\subsection{Time-varying source function}

Instead of choosing a source term that is constant with time, we now implement a source function which is exponentially declining with time, by modifying the injection rate to be
\begin{align}
\dot{C}_\mathrm{inj}(t) = \dot{C}_{\mathrm{inj},0} \e^{-t / \tau_\mathrm{inj}}.
\end{align}
Hence, the injected luminosity is given by
\begin{align}
  L_\mathrm{e}(t) = \dot{C}_\mathrm{inj}(t) \,m_{\e}c^2 A_\mathrm{bol} (p_\mathrm{inj},\alpha).
\end{align}
The energy injected in a timestep $\Delta t_n = t_{n} - t_{n-1}$ is obtained as
\begin{align}
  \Delta E_\mathrm{CRe} &=\int_{t_{n-1}}^{t_n} L_\mathrm{e}(t) \dd t \\
   &= \tau_\mathrm{inj} L_\mathrm{e}(t_{n-1}) \,\e^{-t_{n-1}/\tau_\mathrm{inj}}(1-\e^{-\Delta t/\tau_\mathrm{inj}}).
\end{align}
Since \crest takes the energy density as an input, this has to be divided by the volume to obtain
\begin{align}
  \varepsilon_\mathrm{CRe} = \frac{\Delta E_\mathrm{CRe}}{V_\mathrm{cell}}.
\end{align}

We choose the parameters $\dot{C}_{\mathrm{inj},0}$ and $\tau_\mathrm{inj}$ such that we roughly match the star formation history of our simulated galaxy around 0.5 to 1~Gyr.
In order to do so, we relate the injected luminosity in electrons to the SFR of the galaxy via
\begin{align}
    L_\mathrm{e} = \zeta_\mathrm{SN} \zeta_\mathrm{ep}  \dot{M}_\star \epsilon_\mathrm{SN},
    \label{eq:L_e from SFR}
\end{align}
with the supernova energy released per unit mass $\epsilon_\mathrm{SN} = E_\mathrm{SN}/M_\star = 10^{51}~\mathrm{erg}/(100~\msun) = 10^{49}\mathrm{erg}~\msun^{-1}$. 

This enables us to relate the volume integrated injection rate $\dot{C}_\mathrm{inj}$ to the SFR, $\dot{M}_\star$, via
\begin{align}
    \dot{C}_\mathrm{inj}=\frac{\zeta_\mathrm{SN} \zeta_\mathrm{ep} \epsilon_\mathrm{SN}}{m_\mathrm{e}c^2 A_\mathrm{bol}(p_\mathrm{min}, \alpha_\mathrm{inj})}  \dot{M}_\star.
\label{eq:C_dot_tilde}
\end{align}

\begin{figure*}
    \centering
     \includegraphics[]{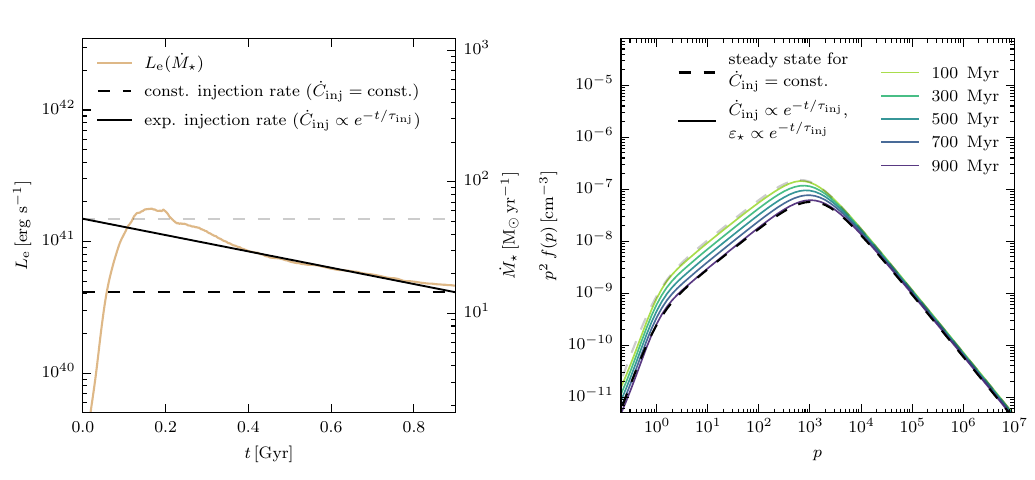}
    \caption{Electron spectrum for an exponentially declining vs. constant source term. 
    \textit{Left-hand panel:} The injected luminosity into electrons $L_\mathrm{e}$ declines, after the initial starburst at 0.2~Gyr, strongly as a function of time in the simulation (brown line). It is derived from the SFR via Eq.~\ref{eq:L_e from SFR}, indicated on the right y-axis. Due to the deviation from a constant source term in the simulation, we set up a one-zone test to compare the effect of an exponentially declining injection rate (solid black line) with two scenarios of a constant injection rate (dashed lines).
    \textit{Right-hand panel:} Evolution of the electron spectra from \crest from a one-zone test with an exponentially declining source term at different times (solid lines). Initially, after 100~Myr, the spectrum builds up and approaches the analytical steady-state spectrum from a constant, high injection rate (dashed gray line), corresponding to the constant, high injection rate shown by a gray dashed line in the left-hand panel. Over time, due to the declining source term, the normalisation decreases (at $p\lesssim2\times10^3$) and the spectrum approaches the steady-state solution with a low, constant injection rate (black dashed line). However, at $p\gtrsim10^4$, where IC cooling dominates, the normalisation stays essentially constant, due to the exponentially declining photon energy density for IC cooling.}
    \label{fig:steady-state-injRate}
\end{figure*}

In our test setup, we implement an exponentially decreasing source term and combine it with an exponentially declining cooling rate for IC cooling.
In particular, we adopt the values $n_\mathrm{gas}=0.24\,\mathrm{cm^{-3}}$ and $B=3.3\mu\mathrm{G}$.
We let the  photon energy density decrease over time as
\begin{align}
    \varepsilon_\star(t) = \varepsilon_{\star,0} \,\e^{-t/\tau_\mathrm{inj}}.
\end{align}
This is motivated by the fact that in our simulation, the incident radiation field for IC cooling is calculated from the SFR, which declines over time.
We start with an initial photon energy density $\varepsilon_{\star, 0}=25\,\varepsilon_\mathrm{CMB}$, which declines to $\varepsilon_\star(t=900~\mathrm{Myr})\approx7\,\varepsilon_\mathrm{CMB}$.

In the left-hand panel of Fig.~\ref{fig:steady-state-injRate}, we show the injection rate into electrons derived from the star formation history of the simulation (using Eq.~\ref{eq:L_e from SFR}) as a function of time. It roughly follows an exponentially declining SFR between 0.4 and 0.9~Gyr, as shown by the black solid line, with $\dot{\tilde{C}}_\mathrm{inj}=5\times10^{-22}\,\mathrm{s^{-1}}\,\mathrm{cm^{-3}}$ at $t=0$,
$\tau_\mathrm{inj}=0.9$~Gyr and distributed over a total volume of a galactic disk with a radius of 20~kpc and a height of 2~kpc. 

To test the evolution of the electron spectra, we input the tracer data created from an exponentially declining source term and the gas properties given above into \crest. The right-hand panel shows the resulting evolution of the electron spectra, color-coded by time. 
We compare the results with two analytical steady-state spectra (using Eq.~\ref{equ:f_steady} with $\alpha=2.2$ and $p_\mathrm{min}=10^{-2}$), with two different values for a constant injection rate: 
A high rate of $\dot{\tilde{C}}_\mathrm{inj}=5\times10^{-22}\,\mathrm{s^{-1}}\,\mathrm{cm^{-3}}$ (gray dashed line) and a lower injection rate of $\dot{\tilde{C}}_\mathrm{inj}=1.4 \times 10^{-23}\,\mathrm{s^{-1}}\,\mathrm{cm^{-3}}$ (black dashed line). Those rates correspond to the initial and final values of the exponentially declining source function (see left-hand panel of Fig.~\ref{fig:steady-state-injRate}), respectively.
Initially, the \crest spectrum builds up towards a steady-state like spectrum similar to the analytical solution with the high injection rate (gray dashed line) at low and high momenta. Over time, it develops the steady-state like spectral shape, but with a decreasing normalisation due to the decreasing source term. Finally, at $t=900$~Myr, it approaches the expected analytical steady-state spectrum from a constant, low injection rate (black dashed line). 

At momenta $p\gtrsim10^3$, the effect of a decreasing normalisation expected from a decreasing source term is almost completely canceled by the diminishing IC cooling over time. This is because at all times, the photon energy density is larger than the magnetic energy density, and hence, IC cooling is the dominant process at high electron momenta. In contrast, at low momenta, Coulomb cooling (which was kept constant with the assumed constant gas density) dominates and hence, the decrease in normalisation is still apparent.
This explains the trends we found in the time evolution of the global electron spectra of our simulation in Fig.~\ref{fig:spectra_times}.

\section{Convergence with resolution}\label{App:Resolution}

\begin{figure}
    \centering
    \includegraphics[]{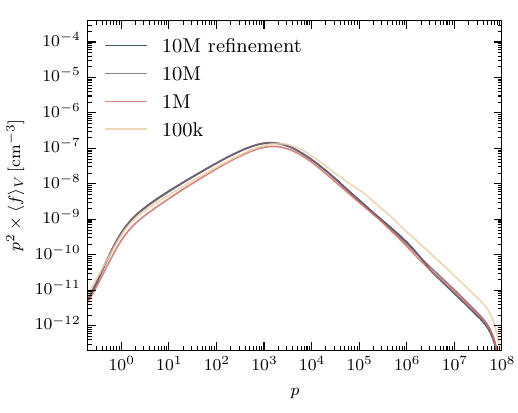}
    \caption{CR electron spectra calculated with \crest on different \arepo simulations, with varying resolution. The spectra are averaged over the galactic disk with a radius of 15~kpc and $\lvert z\rvert < 1$~kpc.}
    \label{fig:CRspectrum_resolutions}
\end{figure}

In Fig.~\ref{fig:CRspectrum_resolutions}, we show the total CR electron spectrum in the galactic disk of four simulations with varying resolution. Clearly, the spectral shapes are well converged below momenta of $10^4$. At larger momenta, we see a more significant deviation for our lowest resolution run (`100k'). All higher resolution runs agree well within the temporal variation we expect on short timescales (see also lower-right panel of Fig.~\ref{fig:spectra_AMS} for the typical range of variation with time).

\begin{figure*}
    \centering
    \includegraphics[]{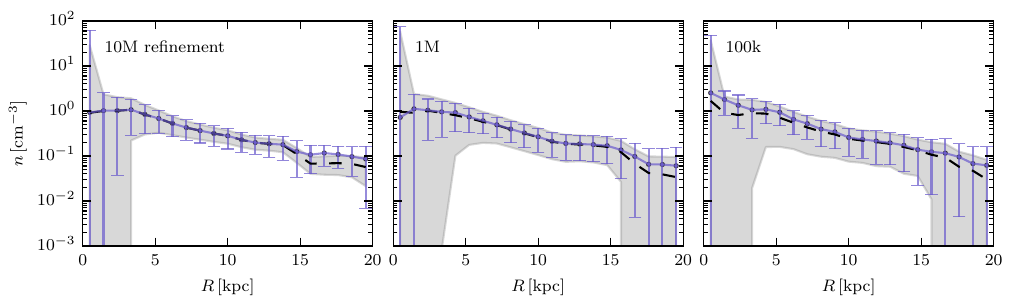}
    \caption{Mean gas density profiles in the \arepo simulations with different resolutions (dashed lines) in 20 cylindrical bins. The shaded regions show the standard deviation of each bin. The tracer particles match the gas density well overall (blue points), with small deviations apparent at large radii $R>15$~kpc, as well as at smaller radii at the lowest resolution.}
    \label{fig:densityprofiles_resolutions}
\end{figure*}

As discussed in Section~\ref{Sec:Simulations}, the velocity tracers used in the simulation to track the necessary quantities for \crest to evolve the CR electron spectra in post-processing are inherently not able to follow the gas flow perfectly. Hence, we typically take a simulation snapshot $\sim$300--500~Myr before the time of interest, place one tracer particle per gas cell in the disk and then continue running the simulation including the tracers. In that way, we manage to trace the gas flow much better, since we avoid the initial collapse of the gas cloud from the very beginning of the simulation, which the tracers do not follow well enough. 

In Fig.~\ref{fig:densityprofiles_resolutions}, we show the gas density profiles at three different resolutions. While in the simulation with $10^5$ gas cells (`100k') the gas density recorded by the tracer particles is on average slightly larger in the central 5~kpc, it is well matched in the simulation with $10^6$ (`10M') and $10^7$ cells (`10M\_refinement') up to radii of 15~kpc. At larger radii, where the gas density is dominated by flows outside the disc, the tracers slightly over-estimate the true density, albeit with a smaller standard deviation in the `10M\_refinement' simulation.


\bsp	
\label{lastpage}
\end{document}